\documentclass[prl,aps,twocolumn,showpacs,superscriptaddress]{revtex4-1}

\usepackage{graphicx,epsfig}
\usepackage{bm}
\usepackage{amsmath}
\usepackage{amssymb}
\usepackage{color}

\usepackage{braket}
\usepackage[backref=true,
bookmarksnumbered=true,
bookmarks=true,
bookmarksopen=true,
colorlinks=true,
citecolor=blue,
linkcolor=blue,
anchorcolor=green,
urlcolor=blue]{hyperref}

\renewcommand{\v}[1]{\ensuremath{\mathbf{#1}}} 
 
\newcommand{\abs}[1]{\left| #1 \right|} 
 
\let\baraccent=\= 
\renewcommand{\=}[1]{\stackrel{#1}{=}} 

\DeclareMathOperator{\Tr}{Tr} 
\DeclareMathOperator{\diag}{\rm diag}

\begin{document}

\title{Robust Majorana Conductance Peaks for a Superconducting Lead}
\smallskip

\author{Yang Peng}
\affiliation{\mbox{Dahlem Center for Complex Quantum Systems and Fachbereich Physik, Freie Universit{\"a}t Berlin, 14195 Berlin, Germany} }

\author{Falko Pientka}
\affiliation{\mbox{Dahlem Center for Complex Quantum Systems and Fachbereich Physik, Freie Universit{\"a}t Berlin, 14195 Berlin, Germany} }

\author{Yuval Vinkler-Aviv}
\affiliation{\mbox{Dahlem Center for Complex Quantum Systems and Fachbereich Physik, Freie Universit{\"a}t Berlin, 14195 Berlin, Germany} }

\author{Leonid I.\ Glazman}
\affiliation{Department of Physics, Yale University, New Haven, CT 06520, USA}

\author{Felix von Oppen}
\affiliation{\mbox{Dahlem Center for Complex Quantum Systems and Fachbereich Physik, Freie Universit{\"a}t Berlin, 14195 Berlin, Germany} }

\begin{abstract}
Experimental evidence for Majorana bound states largely relies on measurements of the tunneling conductance. While the conductance into a Majorana state is in principle quantized to $2e^2/h$, observation of this quantization has been elusive, presumably due to temperature broadening in the normal-metal lead. Here, we propose to use a superconducting lead instead, whose gap strongly suppresses thermal excitations. For a wide range of tunneling strengths and temperatures, a Majorana state is then signaled by symmetric conductance peaks at $eV=\pm\Delta$ of a universal height $G=(4-\pi)2e^2/h$. For a superconducting scanning tunneling microscope tip, Majorana states appear as spatial conductance plateaus while the conductance varies with the local wavefunction for trivial Andreev bound states. We discuss effects of nonresonant (bulk) Andreev reflections and quasiparticle poisoning.
\end{abstract}

\maketitle

{\em Introduction.---}Motivated by possible applications in quantum information processing \cite{Kitaev2003,Nayak2008}, topological superconductors hosting Majorana bound states are currently under intense investigation \cite{review1,review2,review3}. Based on the superconducting proximity effect, various realistic platforms have been proposed to support Majorana states including topological insulators \cite{Fu2008,Fu2009}, semiconductor nanowires \cite{Lutchyn2010,Oreg2010}, and atomic chains \cite{Nadj-Perge2013,Braunecker2013,Vazifeh2013,Klinovaja2013,Pientka2013,Kim2014,Peng2015}. Although these systems are available in the laboratory, the experimental observation of unique Majorana signatures remains challenging.

A widely employed diagnostic tool is the tunneling conductance of normal metal--superconductor junctions, in which Majoranas manifest themselves as characteristic zero-bias peaks \cite{Law2009,Flensberg2010}. Experimental signatures consistent with theoretical predictions have been observed in quantum wires \cite{Mourik2012,Churchill2013,Das2012} and atomic chains \cite{Nadj-Perge2014,Ruby2015b}. However, it is a major challenge in these experiments to uniquely distinguish Majoranas from conventional fermionic subgap states. Spin-polarized subgap states such as Shiba states bound to magnetic impurities \cite{yu,shiba,rusinov,Balatsky2006} or Andreev bound states in a magnetic field can exhibit a zero-energy crossing as a function of exchange interaction or Zeeman energy \cite{Franke2011,Deacon2010,Lee2014}. Thus, such fermionic states may accidentally occur at zero energy and give rise to similar conductance features. As magnetic impurities or external magnetic fields are also required for the most relevant realizations of topological superconductors, such trivial conductance peaks can generally not be disregarded.

In contrast to fermionic subgap states, Majoranas exhibit a celebrated quantized zero-bias conductance of $2e^2/h$ \cite{Law2009,Flensberg2010,Wimmer2011}. Unfortunately, this has so far proved difficult to observe in experiment. The Fermi distribution in the metal lead is smooth on the scale of the temperature $T$, which strongly limits the experimental energy resolution. When temperature is larger than the tunnel coupling, the Majorana peak is broadened and the zero-bias conductance is reduced. Even at low temperatures (e.g., $T=60\,$mK in Ref.~\cite{Mourik2012}), it may be difficult to observe the quantized peak height as multichannel effects limit the relevant tunneling strength \cite{Pientka2012}. Quasiparticle poisoning may also lead to deviations from quantization. A fermion-parity breaking rate exceeding the tunnel coupling broadens the peak and reduces its height. This requires one to work at temperatures below the lowest fermionic excitations in the topological superconductor.

In this paper, we show how a robust conductance signature of Majorana bound states can be obtained by employing superconducting leads. In striking contrast to normal-state contacts, effects of thermal broadening are strongly suppressed for a superconducting lead because quasiparticle excitations are exponentially suppressed $\sim \exp(-\Delta/T)$ by its superconducting gap $\Delta$. Majorana bound states no longer appear as zero-bias anomalies but rather as two symmetric peaks in the differential conductance $G=dI/dV$ which occur when the BCS singularity of the superconducting gap lines up with the Majorana bound state, i.e., at the thresholds $eV=\pm \Delta$. These peaks have a universal height 
\begin{align}
G_{\rm M}=(4-\pi)\frac{2e^2}{h},\label{majorana_conductance}
\end{align}
which persists over a wide range of tunnel couplings.

This yields particularly striking evidence when employing a scanning tunneling microscope (STM) with a superconducting tip which allows for spatially resolved measurements. This has previously been used to map out bound state wavefunctions in conventional and unconventional superconductors
\cite{Yazdani1997,Yazdani1999,Hudson2001,Ji2008,Nadj-Perge2014,Ruby2015b}. Here we propose that such maps can clearly distinguish between Majoranas and trivial zero-energy bound states. Indeed, the peak conductance is uniform in the vicinity of Majorana states and a conductance map exhibits a characteristic mesa or plateau structure. In contrast, the conductance of trivial subgap states exhibits a spatial pattern which is governed by the bound-state wavefunction. 

In addition, STM measurements allow for systematic studies as a function of tunneling strength by varying the tip height. It was recently demonstrated \cite{Ruby2015a} that this can be exploited to probe quasiparticle relaxation processes. In the present context, varying the tunneling strength may help to identify Majorana signatures despite competing effects such as nonresonant Andreev reflections or quasiparticle poisoning. 

{\em Subgap conductance for Majorana bound state.---}At subgap voltages $eV<\Delta+\Delta_s$ and zero temperature, the tunneling current between  superconducting tip or lead and substrate (with gap $\Delta_s$) flows by multiple Andreev reflections. Near the threshold $e|V|=\Delta$, the differential conductance ${\mathrm d}I/{\mathrm d}V$ is dominated by single Andreev reflections from the sample. For tip locations far from the zero-energy bound state in the sample, this yields the familiar peak in ${\mathrm d}I/{\mathrm d}V$ due to the singular densities of states of incoming electrons and outgoing holes. In the vicinity of the bound state, tunneling is further enhanced by the zero-energy resonance \cite{Badiane2011,SanJose2013,Ruby2015a}.

Formally, the subgap current due to single Andreev reflections from the sample can be expressed as \cite{Cuevas1996,supp,Martin2014} 
\begin{align}
    I&(V)=4e \pi^2t^4\int \frac{d\omega}{2\pi\hbar}{\rm Tr} [ G_{eh}(r,\omega)G_{eh}^\dag(r,\omega)] \nonumber\\
 &\times \rho(\omega_-)\rho(\omega_+)[n_F(\omega_-)-n_F(\omega_+)],\label{current}
\end{align} 
where $t$ is the amplitude for tip-substrate tunneling, $\omega_\pm=\omega\pm eV$, $n_F(\omega)$ denotes the Fermi function, and the superconducting tip enters through its BCS density of states $\rho(\omega)=\nu_0\theta(|\omega|-\Delta)|\omega|/\sqrt{\omega^2- \Delta^2}$ with $\nu_0$ the normal density of states at the Fermi energy. Spin or subband degrees of freedom are accounted for by a possible matrix structure of the anomalous retarded Green function $G_{eh}(r,\omega)$ of the substrate at the tip position $r$. In terms of its Lehmann representation, $G_{eh}(r,\omega)$ has contributions from both the bound state and the above-gap continuum. In the following, we first consider the resonantly enhanced Andreev current from a Majorana bound state and subsequently discuss the contribution of the quasiparticle continuum. 

For $e|V|\simeq \Delta$, we can approximate $n_F(\omega_-)-n_F(\omega_+)\simeq {\rm sgn} V$ in Eq.\ (\ref{current}), up to corrections of order $\exp(-\Delta/T)$. This insensitivity to temperature is a key advantage of superconducting leads. The bound-state contribution to the substrate Green function is 
\begin{equation}
   G(r,\omega)=\frac{\langle r | \psi\rangle \langle \psi | r \rangle} {\omega+i\Gamma/2}.
   \label{Dirac_green_function} 
\end{equation}
Here, $\langle r |\psi\rangle=[\zeta(r),\pm {\Theta}\zeta(r)]^{\rm T}$ denotes the local Bogoliubov--de Gennes wavefunction of the Majorana bound state with ${\Theta}$ the time-reversal operator. The broadening $\Gamma = 2i\braket{\psi|\Sigma|\psi}$ of the bound state is induced by the tunnel coupling to the lead. The corresponding self energy $\Sigma=-i\pi t^2{\rm diag}[\rho(\omega_-),\rho(\omega_+)]$ is diagonal as Andreev reflections in the lead can be neglected near $e|V|=\Delta$.
 
\begin{figure}[t]
\includegraphics[width=0.45\textwidth]{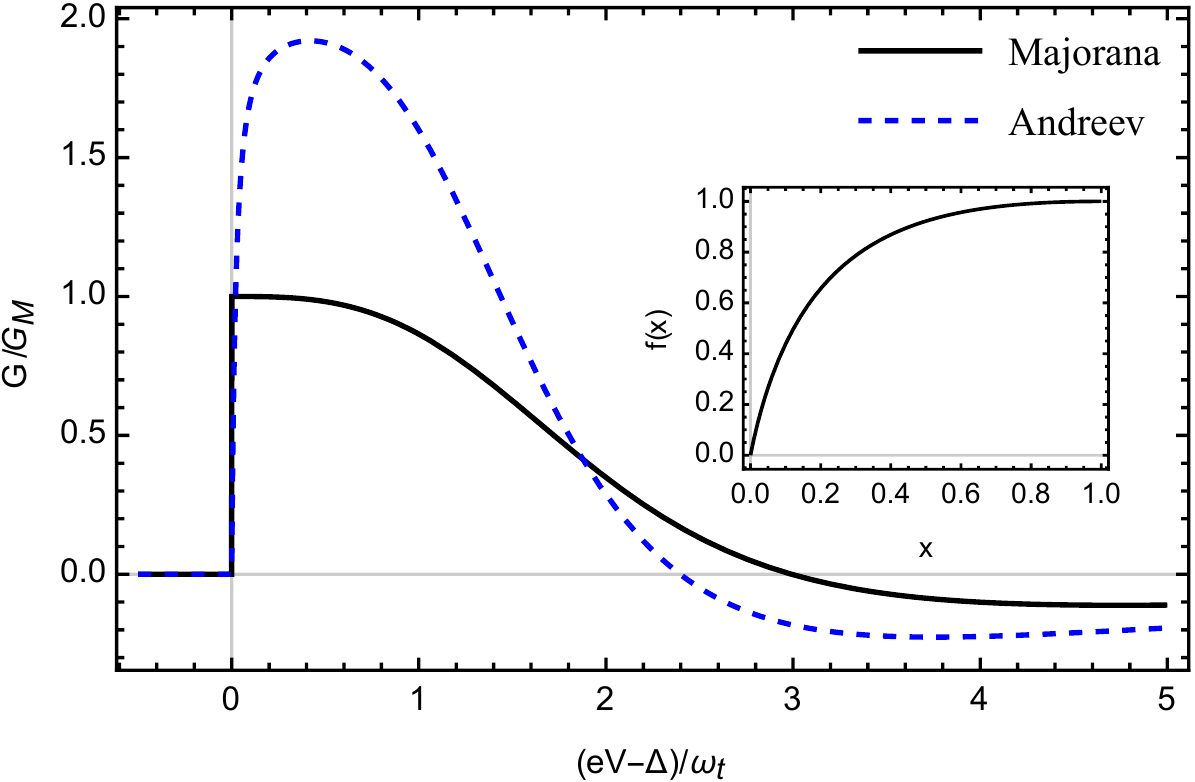}\\
\includegraphics[width=.237\textwidth]{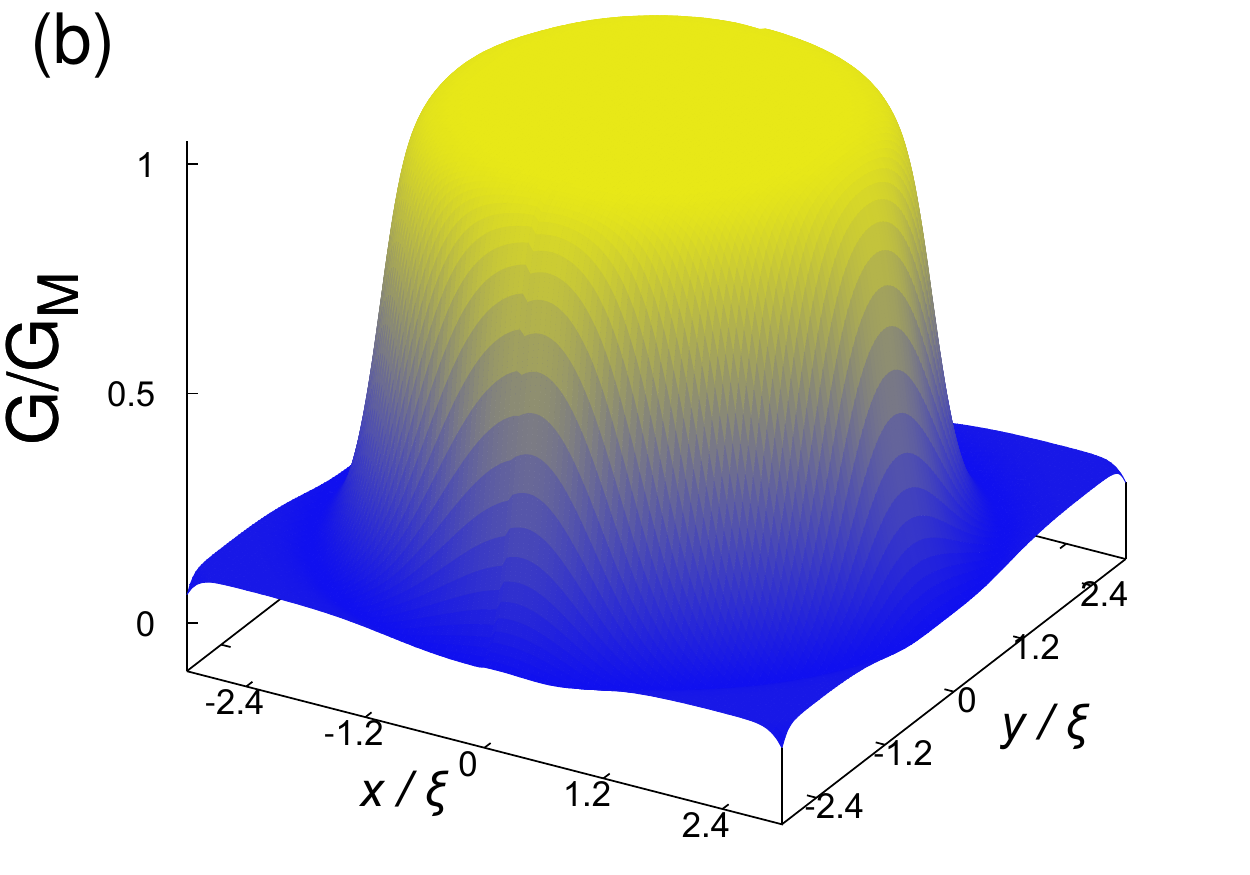}
\includegraphics[width=.237\textwidth]{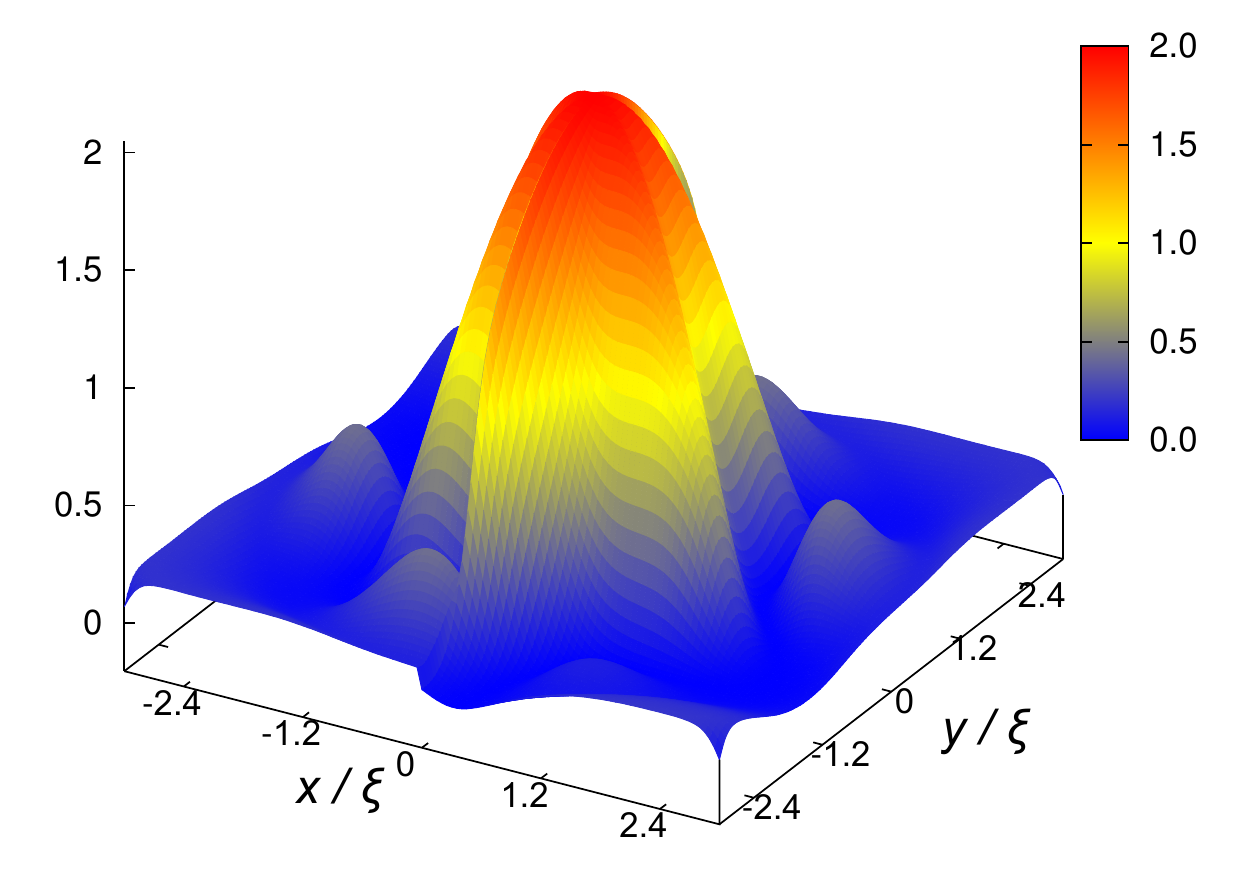}
\caption{(color online) (a) Differential conductance {\em vs} bias voltage near the threshold $eV=\Delta$ for Majorana (solid line) and Andreev state with $|u|=|v|$ (dashed line). For a Majorana, the conductance exhibits a step of height $(4-\pi)2e^2/h$ at the threshold. For an Andreev state, the conductance has a smooth onset, cf.\ Eq.~(\ref{dirac_conductance_bulk}). Both peaks have a negative-differential conductance dip at high voltages.  Inset: Graph of $f(x)$ as defined in the main text. (b) Spatial conductance maps for Majorana (left) and Andreev state (right) for $\omega_t(0)/\delta\Delta=5$. The Majorana gives rise to a conductance plateau whereas the Andreev state exhibits a pattern reflecting the spatial dependence of the ratio $(u/v)^2$. The Majorana conductance drops far from the bound state when the broadening exceeds $\omega_t$.}
\label{fig:lineshape}
\end{figure}

Inserting Eq.\ (\ref{Dirac_green_function}) into (\ref{current}) yields (for $V>0$) \cite{Levy-Yeyati1997,Ruby2015a}
\begin{align}
    I=\frac{e}{h}\int d\omega \frac{\Gamma_e(\omega)\Gamma_h(\omega)}{\omega^2+[\Gamma_e(\omega)+\Gamma_h(\omega)]^2/4} 
\label{diraccurrent}
\end{align}
in terms of the electron and hole tunneling rates $\Gamma_{e/h}(\omega)=2\pi t^2|\zeta|^2\rho(\omega_\mp)$. While the integrand in Eq.\ (\ref{diraccurrent}) has a resonance denominator, its behavior is peculiar due to the strong energy dependence of the tunneling rates. Specifically, the square-root singularity of the BCS density of states implies that the integrand involves a characteristic energy scale $\omega_t=(\pi t^2\nu_0 |\zeta(r)|^2\sqrt{\Delta/2})^{2/3}$ which depends on a fractional power of the tunneling rate from a normal tip $\gamma_n=2\pi t^2\nu_0 |\zeta(r)|^2$. In the weak-tunneling regime $\omega_t\ll\Delta$, we can write
\begin{equation}
   I=\frac{4e}{h} \int_{-\eta}^{\eta}\frac{d\omega}{\sqrt{\eta^2-\omega^2}}\frac{\omega_t^3} {\omega^2+\omega_t^{3} 
   \bigl(\frac{1}{\sqrt{\eta-\omega}}+\frac{1}{\sqrt{\eta+\omega}}\bigr)^2},
   \label{threshold_current}
\end{equation}
for $0<\eta\ll\Delta$, where $\eta=eV-\Delta$ measures the voltage from the threshold $\Delta$. In the vicinity of the threshold, $\eta \ll \omega_t$, the resonance denominator is dominated by the second term and we obtain $I(V) = (4-\pi)(2e/h)(eV-\Delta)\theta(eV-\Delta)$ and thus Eq.~(\ref{majorana_conductance}). The entire peak lineshape
\begin{equation}
   \frac{{\mathrm d}I}{{\mathrm d}V} = (4-\pi)\frac{2e}{h}\,\Lambda\!\left(\frac{eV-\Delta}{\omega_t}\right),
\label{current_at_Delta}
\end{equation}
involves the function $\Lambda(x)$ which vanishes for $x<0$, jumps to $\Lambda(0^+) = 1$, and falls off with a small negative differential conductance tail at large $x$, cp.\ Fig.\ \ref{fig:lineshape}. 

Thus, the differential conductance between a conventional superconductor and a Majorana state exhibits a peak which is independent of tunneling strength and Majorana wavefunction. While the peak height is close to the quantized Majorana peak height $2e^2/h$ for a normal-metal lead, there are several differences: (a) There are two symmetric, finite-bias Majorana peaks at $eV=\pm\Delta$ rather than a single zero-bias peak, (b) the conductance peak is strongly asymmetric with a discontinuous step at the threshold, and (c) the width of the peak is set by $\omega_t$ with its sublinear dependence on junction transparency.

The threshold discontinuity in the conductance persists even when including the contributions of the quasiparticle continuum in the substrate Green function. To see this, we model the substrate superconductor by a $2\times2$ Nambu Green function $g(\omega,r)$. For a topological substrate, this is appropriate for perfect spin polarization (spinless $p$-wave superconductor). Including the tunnel coupling to the tip through the self energy $\Sigma$ as given above, the substrate Green function becomes $G=g[1-\Sigma g]^{-1}$. We first focus on the vicinity of the bound state where the conductance is dominated by Andreev reflections from the bound state. By straightforward calculation and expansion of $g$ in $\omega$ \cite{supp}, we find
\begin{equation}
   G(r,\omega)=\frac{\langle r|\psi\rangle \langle\psi| r\rangle}{\omega-\lambda(\omega)+i\Gamma/2} \label{bulk_green_function}.
\end{equation}
This differs from the pure bound-state contribution by the additional term $\lambda(\omega)=\pi^2 t^4\omega\det g(\omega,r)\rho(\omega_-) \rho(\omega_+)$ in the denominator which involves the determinant (in particle-hole space) of the bare substrate Green function. While the determinant of the bound-state contribution to the Green function vanishes, this is no longer the case when including the quasiparticle continuum. At subgap energies away from bound states, the Green function $g(\omega,r)$ is a hermitian $2\times2$ matrix, so that $\det g(\omega,r)$ and hence $\lambda(\omega)$ are real. Thus, we find
\begin{align}
 I_M(V)=\frac{4e}{h} 
 \int_{-\eta}^{\eta}\frac{d\omega}{\sqrt{\eta^2-\omega^2}}\frac{\omega_t^3}{(\omega-\lambda)^2+\bigl(\frac{\omega_t^{3/2}}{\sqrt{\eta-\omega}}
 +\frac{\omega_t^{3/2}}{\sqrt{\eta+\omega}}\bigr)^2}.
 \label{threshold_current_bulk}
\end{align}
For a Majorana state, the real part of the resonance denominator must vanish exactly at $\omega=0$. Indeed, particle-hole symmetry further constrains $\det g(\omega,r)$ to be an even function of $\omega$ which can be approximated as a constant at small $\omega$ (see \cite{supp}, where this conclusion is confirmed by model calculations). Then, we find $\lambda(\omega)\propto t^4 \omega/\sqrt{\eta^2-\omega^2}$ near the threshold. Even with this term, the denominator in Eq.~(\ref{threshold_current_bulk}) remains dominated by the divergent tunnel broadenings $\sim\omega_t^{3/2}/\sqrt{\eta\pm \omega}$ and the discontinuous conductance step as well as the universal value of the threshold conductance in Eq.~(\ref{majorana_conductance}) persist.

In experiment, the square-root singularity of the BCS density of states of the tip may be broadened intrinsically due to higher-order processes or effectively due to experimental resolution. The universal threshold conductance persists as long as $\omega_t$ exceeds this broadening.
This condition also determines the spatial extent of the conductance plateau, $r\lesssim 4\xi \ln[\omega_t(0)/\delta\Delta]/3$, where $\xi$ is the Majorana localization length, $\omega_t(0)$ denotes the value of $\omega_t$ at the center of the Majorana bound state, and $\delta\Delta$ is the broadening of the tip density of state, cf.\ Fig.\ \ref{fig:lineshape}(b). Of course, a well-resolved Majorana peak also requires $\omega_{t}\ll \Delta_s$, i.e., the tunnel broadening needs to be small compared to the induced gap. If the peak is not fully resolved, it is suppressed below the universal value and its height may vary as a function of space. 

For tip locations far from the bound state, the tunneling conductance is dominated by conventional (```nonresonant") Andreev reflections. These still yield a threshold peak due to the singular tip density of states in $\Gamma_e$ and $\Gamma_h$, but are not enhanced by a bound-state resonance. For a one-dimensional $p$-wave superconductor, this conductance peak has height $\simeq 1.3G_M$ and width $\sim \Delta {\cal T}^2$ quadratic in the junction transparency ${\cal T}\propto t^2$  \cite{supp}. Observing the conventional Andreev peak thus requires that the broadening of the tip density of states is small compared to $\sim \Delta {\cal T}^2$. This is a much more stringent condition than for the resonant Andreev peak as the width of the bound-state peak $\omega_t\propto t^{4/3}$ involves a lower power of $t$. We note that in a typical STM experiment \cite{Ruby2015a}, conventional Andreev peaks can be resolved only for small tip-sample distances, while bound-state signatures persist to much weaker tunnel couplings.

{\em Subgap conductance for Andreev bound state.---}These results should be contrasted with those for trivial zero-energy Andreev bound states. For concreteness, consider an $s$-wave superconductor with conserved spin \cite{foot1}, whose Bogoliubov--de Gennes description decomposes into two independent spin sectors that interchange under particle-hole transformations. A zero-energy Andreev state corresponds to two Bogoliubov--de Gennes wavefunctions, $\langle r|\psi_+\rangle =[u(r),v(r)]^{\rm T}$ and $\langle  r|\psi_-\rangle =[\Theta v(r),-\Theta u(r)]^{\rm T}$, one in each sector. An analogous calculation \cite{supp} yields the threshold current
\begin{align}
   I_A(V)=2I_M(V)f(|u(r)|^2/|v(r)|^2). \label{dirac_current}
\end{align}
Reflecting the two zero-energy wavefunctions, the maximal threshold conductance is twice that in the Majorana case, $G_A=2G_M$, and realized for the particle-hole symmetric case $|u|=|v|$. In general, the peak conductance depends on the ratio of electron and hole wavefunction at the tip position. This dependence is captured by the dimensionless function $f(x)=\frac{2x}{4-\pi}\int_{-1}^1 dz \sqrt{1-z^2}/(x\sqrt{1-z}+\sqrt{1+z})^2$ which takes on values between 0 and 1 and is plotted in Fig.~\ref{fig:lineshape}(a). The function satisfies $f(x)=f(1/x)$ as the two spin sectors contribute equally.
In the limit of large particle-hole asymmetry, $G_A\sim G_M {\rm min}(|u/v|^2, |v/u|^2)\ll G_M$. The lineshape of the conductance peak is similar to the Majorana peak, with a width of order $\omega_t$ upon replacing $\zeta(r)$ by ${\rm max}\{u(r),v(r)\}$. 

Our results imply that the height of the conductance peak allows for a clear distinction between a conventional Andreev bound state and a Majorana state. Even when $f(u^2/v^2)\sim 1/2$ for one location of the STM tip, moving the tip to another location modifies the conductance peak height for a conventional bound state, tracking the ratio of electron and hole wavefunctions. In contrast, the conductance map exhibits a characteristic mesa structure for a Majorana state, see Fig.~\ref{fig:lineshape}(b). In non-STM tunneling experiments, changes of parameters (e.g., gate voltages) which affect the Majorana wavefunction should leave the peak height unchanged for a Majorana but not for a conventional Andreev bound state. 

As there is no locking of the bound state to zero energy, also the continuum contribution is distinctly different for conventional Andreev states. The two spin sectors are described by separate $2\times2$ Nambu Green functions which map into one another under particle-hole transformations. This is quite unlike the Majorana Green function which maps onto itself. For each sector, $\det g(\omega,r)$ is therefore no longer an even function of $\omega$ and will generally have a singular contribution $\propto 1/\omega$ at the threshold so that $\lambda(\omega)\sim {\cal T}^2\Delta_s\Delta/ \sqrt{\eta^2-\omega^2}$. These general arguments can be confirmed explicitly for Shiba states in $s$-wave superconductors \cite{supp}. Near the threshold, the resonance denominator in the expression for the current is now dominated by $\lambda(\omega)$. As illustrated in Fig.~\ref{fig:lineshape} by a numerical evaluation of the current, this suppresses the conductance step. Analytically, we find that just above the threshold, the conductance increases linearly,
\begin{align}
 G_A(V)\sim \frac{2e^2}{h}\frac{1}{{\cal T}^2}\frac{eV-\Delta}{\Delta}\theta(eV-\Delta),\label{dirac_conductance_bulk}
\end{align} 
and matches with the conductance obtained from Eq.~(\ref{dirac_current}) for $eV-\Delta\gg {\cal T}^2\Delta$. We note that this suppression of the conductance step depends on ${\cal T}$ and can thus be probed by varying the tip-sample distance in an STM experiment. This may serve as an additional signature to distinguish between Majorana and conventional Andreev bound states. 

\begin{figure}[t]
\includegraphics[width=0.45\textwidth]{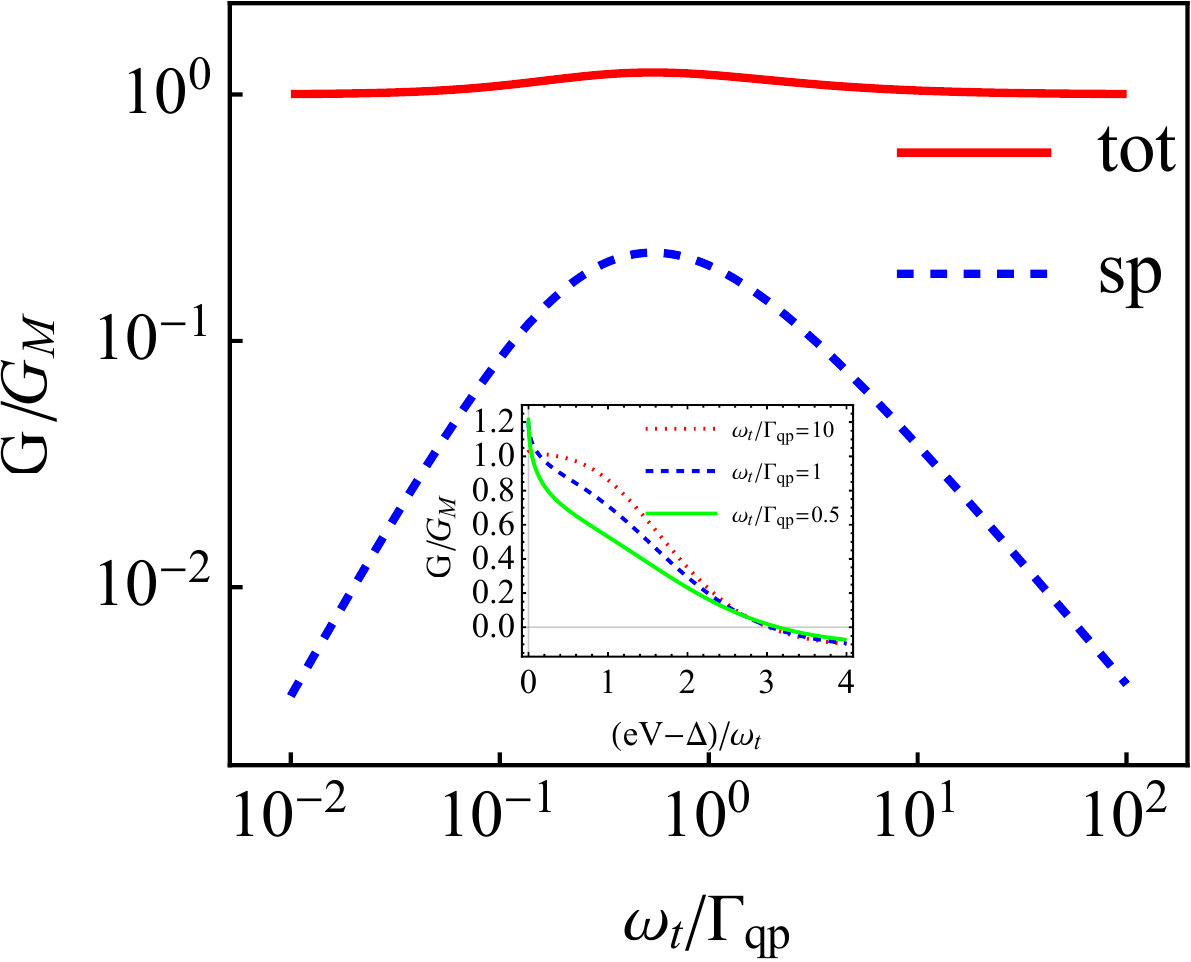}
\vspace{0pt}
\caption{Total threshold conductance for a Majorana state (tot) along with the single-particle contribution (sp) as a function of $\omega_t$. The single-particle contribution affects the conductance only a in a window of transmission values, where $\omega_t \sim \Gamma_{\rm qp}$. While the maximum is of order $0.2 G_M$, the position of the maximum in tunneling strength depends sensitively on temperature (through $\Gamma_{\rm qp}$). Inset: Line shape of the total conductance as a function of voltage away from the threshold, for different ratios of $\omega_{t}/\Gamma_{qp}$.}
\label{fig:peakheight}
\end{figure}

{\em Effects of quasiparticle poisoning.---}So far, we only included bound-state broadening by the tunneling contact. At finite temperatures, the bound-state occupation also changes by inelastic transitions to other subgap states or the quasiparticle continuum in the sample \cite{poisoning}. We account for these processes by an additional contribution $i\Gamma_{\rm qp}/2$ to the self energy of the bound-state Green function Eq.~(\ref{bulk_green_function}). This does not affect the Andreev current at the threshold, where the denominator is dominated by the diverging tunnel coupling. However, the overall weight of the peak is reduced by a {\em narrowing} of the linewidth by a factor $(\omega_t/\Gamma_{\rm qp})^2$ once $\Gamma_{\rm qp}>\omega_t$, see Fig.\ \ref{fig:peakheight} (inset). 

In addition, quasiparticle poisoning generates a single-electron current $I^s$ which involves tunneling of single particles followed by inelastic transitions from the zero-energy
bound state to other bound states or the quasiparticle continuum \cite{Ruby2015a}. For a Majorana state, we find near the threshold $eV=\Delta$ (with analogous results applying for Andreev bound states) \cite{supp}
\begin{align}
 I^s_M=&\, \frac{e}{4h}\int d\omega  \frac{\Gamma_{\rm qp} [\Gamma_e(\omega) +\Gamma_h(\omega)]}{\omega^2+[\Gamma_{\rm qp} +\Gamma_e(\omega)+\Gamma_h(\omega)]^2/4}.
\end{align} 
For weak and strong tunneling, this yields \cite{supp}
\begin{align}
G^s_M \sim \frac{2e^2}{ h}\left\{ \begin{array}{ll}
\displaystyle\left(\omega_t/\Gamma_{\rm qp}\right)^{3/2}  \quad &\omega_t\ll\Gamma_{\rm qp},\\
\displaystyle \Gamma_{\rm qp}/\omega_t \quad &\omega_t\gg\Gamma_{\rm qp}.
\end{array}\right.
\end{align}
Figure~\ref{fig:peakheight} shows that this single-particle contribution assumes a maximum of $\sim 0.2G_M$ when $\omega_t\sim \Gamma_{\rm qp}$. However, it can be easily made negligible by tuning the system away from this maximum through varying temperature or tunneling strength.

{\em Conclusions.---}We show that conductance measurements with superconducting leads constitute a promising technique to identify Majorana states. The presence of Majoranas is signaled by conductance peaks of universal height which are largely unaffected by thermal broadening, a key obstacle in previous experiments with normal-metal contacts. We discuss strategies to systematically rule out parasitic effects such as quasiparticle poisoning or trivial subgap states. The proposed setup is readily available in the laboratory and, in fact, has already been realized in previous experiments \cite{Nadj-Perge2014,Ruby2015b,Deng2012,Finck2013}. (Notice, however, that temperature was comparable to the induced gap in the STM experiments performed to date, precluding observation of the universal conductance, and that the nanowire experiments focused on zero-bias peaks.) Our results also imply that quasiparticle poisoning rates can be extracted from systematic measurements as a function of tip height and temperature. 

{\em Acknowledgments.---}We thank P.\ Brouwer, K.\ Franke, B.\ Heinrich, J.\ Meyer, Y.\ Oreg, and M.-T.\ Rieder for stimulating discussions. We acknowledge financial support by the Helmholtz Virtual Institute ``New states of matter and their excitations," SFB 658, SPP1285 and SPP1666 of the Deutsche Forschungsgemeinschaft, the Humboldt Foundation, the Minerva Stiftung, as well as DOE contract DE-FG02-08ER46482 at Yale University. We are grateful to the Aspen Center for Physics, supported by NSF Grant No.\ PHYS-106629, for hospitality while this line of work was initiated.

\clearpage

\onecolumngrid

\section*{Supplementary material}

\section{General formula for Andreev current into subgap states}

In this section, we outline the derivation of the tunneling current in Eq.~(2) of the main text. This standard calculation is included to make the presentation self contained and closely follows the derivation for $2\times 2$ Green functions presented in Ref.\ \cite{Ruby2015a}. We describe the tunneling contact by the Hamiltonian $\hat{H}=\hat{H}_{L}+\hat{H}_{R}+\hat{H}_{T}$, where the three terms refer to the lead (tip), the sample, and the tunnel coupling. The superconducting tip with chemical potential $\mu$ and gap $\Delta$ is described by the BCS Hamiltonian 
\begin{align}
 \hat{H}_{L}=\int \frac{d\mathbf{k}}{(2\pi)^3} \biggl[\sum_\sigma \xi_k   \hat{c}_{L,\mathbf{k}\sigma}^\dag  \hat{c}_{L,\mathbf{k}\sigma} +(\Delta \hat{c}_{L,\mathbf{k}\uparrow}^\dag \hat{c}_{L,-\mathbf{k}\downarrow}^\dag +{\rm h.c.})\biggr],
\end{align}
where $\xi_k=k^2/2m-\mu$ and $c_{L,\mathbf{k}\sigma}$ ($c^{\dagger}_{L,\mathbf{k}\sigma}$) annihilates (creates) an electron in the tip with momentum $\mathbf{k}$ and spin $\sigma$.
The sample Hamiltonian generally takes the form
\begin{align}
\hat{H}_{R}=\int dx\, \sum_{\sigma,\sigma'} \hat{c}_{R,\sigma}^\dagger(x) \mathcal{H}_{R,\sigma\sigma'}(x) \hat{c}_{R,\sigma'}(x),
\end{align}
where $\mathcal{H}_{R,\sigma\sigma'}(x)$ is the Hamiltonian in first quantization and $\hat{c}_{R,\sigma}(x)$ annihilates an electron with spin $\sigma$ at position $x$ in the sample. We choose the superconducting order parameters in tip and sample to be real such that the superconducting phase difference $\phi(\tau)$ enters the tunneling Hamiltonian
\begin{equation}
\hat{H}_{T}(\tau)=\sum_{\sigma}\left[te^{i\phi(\tau)/2}\hat{c}_{L,\sigma}^{\dagger}(0,\tau)\hat{c}_{R,\sigma}(x,\tau)+te^{-i\phi(\tau)/2}\hat{c}_{R,\sigma}^{\dagger}(x,\tau)\hat{c}_{L,\sigma}(0,\tau)\right],
\end{equation}
where $\tau$ the time argument, $t$ is the hopping strength, and $\hat{c}_{L,\sigma}(0,\tau)=\int d\mathbf{k}\hat{c}_{L,\mathbf{k}\sigma}(\tau)/(2\pi)^3$ annihilates an electron in the tip at the tunneling contact, which is located at the origin. The sample is contacted at position $x$ and we suppress the position arguments in the following for simplicity. The time-dependent phase difference between the tip and the sample, $\phi(\tau)=\phi_{0}+2eV\tau$, depends on the voltage $V$ applied to the junction.

We evaluate the current from the Heisenberg equation of motion $\hat{I}=-e\dot{\hat{N}}_{L}=ie[\hat{N}_{L},\hat{H}_{T}]$, where $\hat{N}_{L}$ is the electron-number operator of the tip. Taking the expectation value, we obtain
\begin{equation}
I(\tau)=\frac{e}{2}\Tr\left\{\tau_{z}\left[\hat{t}(\tau)G_{RL}^{<}(\tau,\tau)-G_{LR}^{<}(\tau,\tau)\hat{t}^{*}(\tau)\right]\right\},
\end{equation}
where $\tau_z$ is a  Pauli matrix acting in Nambu space, $\hat{t}(\tau)=te^{i\tau_z\phi(\tau)/2}\tau_z$, and we have introduced the lesser Green function in Nambu and spin space with matrix elements given by $(G^{<}_{\alpha\beta})_{ij} = i\langle \Psi^{\dagger}_{\beta j} \Psi_{\alpha i}\rangle$. Here the $\Psi_{\alpha j}$ are components of the Nambu spinor $\Psi_{\alpha}=\left( c_{\alpha,\uparrow}, c_{\alpha,\downarrow}, c_{\alpha,\downarrow}^{\dagger},-c_{\alpha,\uparrow}^{\dagger}\right)^T$ with $\alpha,\beta=L,R$.

We are interested in tunneling processes to lowest order in the tunneling amplitude and therefore neglect Andreev reflections in the tip, which give rise to multiple Andreev reflections. As these processes involve several single-particle tunneling events they enter only at higher orders in the tunneling amplitude.
This approximation is exact when the sample is spin polarized (e.g., a proximity-coupled semiconductor nanowire in a strong magnetic field) and one of the two spin components is fully normal reflected. In this case, spin-flipping Andreev reflections in the tip do not contribute to transport.

We follow the nonequilibrium Green function approach described in Ref.~\cite{Ruby2015a,Cuevas1996} setting the offdiagonal elements of the tip Green function in Nambu space to zero and denoting the diagonal elements by $g_L(\omega)$. We obtain the $dc$ current 
\begin{equation}
    I =\frac{et^{2}}{2h}\int d\omega\,\Tr\left\{ G_{R}^{>,ee}(\omega)g_{L}^{<}(\omega-eV)-G_{R}^{<,ee}(\omega)g_{L}^{>}(\omega-eV)-g_{L}^{>}(\omega+eV)G_{R}^{<,hh}(\omega)+g_{L}^{<}(\omega+eV)G_{R}^{>,hh}(\omega)\right\}, \label{current_formula_green_function}
\end{equation}
where $G_{R}(\omega)=\int d\tau_1d\tau_2 \exp[i\omega(\tau_1-\tau_2)]G_R(\tau_1,\tau_2)$ is the sample Green function in presence of the tip, $(e,h)$ are indices in Nambu space denoting particle and hole components, and the trace is taken in spin space. This expression has been derived in Ref.~\cite{Ruby2015a} for the case when tip and sample are both spin-conserving $s$-wave superconductors.

According to the Langreth rule, the lesser Green function of the sample can be written as
\begin{align}
 G_R^<&=g_R^<+g_R^r\Sigma_R^rG_R^<+g_R^r\Sigma_R^<G_R^a+g_R^<\Sigma_R^aG_R^a \nonumber \\
 &=\left(1-g_R^r\Sigma_R^r\right)^{-1}g_R^<(1+\Sigma_R^aG_R^a) + G_R^r\Sigma_R^<G_R^a.
 \label{eq:GF_l}
\end{align}
Similar expressions also exist for the greater Green function. The first term involving $g_R^<$ gives rise to a single-particle current which at subgap energies requires inelastic processes in the sample. The second term leads to the current carried by Andreev reflections. The sample self energy due to the presence of the tip can be written as
\begin{equation}
    \Sigma(\omega)=t^{2}\diag(g_{L}(\omega_{-}),g_{L}(\omega_{-}),g_{L}(\omega_{+}),g_{L}(\omega_{+})),\quad \omega_{\pm}=\omega\pm eV \label{eq:self-energy}.
\end{equation}

We now focus on the Andreev current and defer the discussion of the single-particle current to Sec.~\ref{sec:single-particle-current}. Thus neglecting the first term in Eq.~(\ref{eq:GF_l}) we find Eq.~(2) of the main text, 
\begin{align}
I(V) & =\frac{et^{4}}{2h}\int d\omega\,\Tr\left\{ G_{R}^{r,eh}(\omega)g_{L}^{>}(\omega+eV)G_{R}^{a,he}(\omega)g_{L}^{<}(\omega-eV)-G_{R}^{r,eh}(\omega)g_{L}^{<}(\omega+eV)G_{R}^{a,he}(\omega)g_{L}^{>}(\omega-eV)\right\} \nonumber \\
 & +\frac{et^{4}}{2h}\int d\omega\,\Tr\left\{ g_{L}^{>}(\omega+eV)G_{R}^{r,he}(\omega)g_{L}^{<}(\omega-eV)G_{R}^{a,eh}(\omega)-g_{L}^{<}(\omega+eV)G_{R}^{r,he}(\omega)g_{L}^{>}(\omega-eV)G_{R}^{a,eh}(\omega)\right\} \nonumber \\
 & =\frac{4\pi^{2}et^{4}}{h}\int d\omega\,\|G_{R}^{eh}(\omega)\|^{2}\rho(\omega+eV)\rho(\omega-eV)\left[n_{F}(\omega-eV)-n_{F}(\omega+eV)\right]
 \label{eq:current}
\end{align}
where $G_{R}^{r,eh}$ is the electron-hole block of the (retarded) Green function of the sample and $\|G\|=\sqrt{\Tr\left(GG^{\dagger}\right)}$
denotes the Frobenius norm of matrix $G$. Here we have used the relations $g_{L}^{<}(\omega)=2\pi in_{F}(\omega)\rho(\omega)$ and $g_{L}^{>}(\omega)=-2\pi i(1-n_{F}(\omega))\rho(\omega)$, where $n_{F}(\omega)$ is the
Fermi distribution function, $\rho(\omega)=\nu_{0}\abs{\omega}\theta(\omega^{2}-\Delta^{2})/\sqrt{\omega^{2}-\Delta^{2}}$ and $\nu_{0}$ is the normal density of state at the Fermi energy in the tip. Due to the step functions in $\rho(\omega\pm eV)$ the integration interval is restricted to $\omega\in(-(eV-\Delta),eV-\Delta)$. Note that in this interval the self energy is purely imaginary.

\section{Conductance for zero-energy bound states}\label{sec:bound_state_conductance}

In this section, we calculate the conductance for isolated Majorana or Andreev states at zero energy, cf.\ Eqs.~(6) and (9) of the main text, neglecting the contributions of all other states in the sample. While the main text focuses on Andreev states in $s$-wave superconductors, we also consider more general spin structures here.

\subsection{Topological superconductor with Majorana bound states}

We first consider a topological superconductor substrate with a single zero-energy Majorana state to the tip. The Majorana wavefunction has the form
$\Phi_{0}(x)=\left(u_{\uparrow}(x),u_{\downarrow}(x)^{*},u_{\downarrow}(x),-u_{\uparrow}(x)^{*}\right)^{T}$ which maps onto itself under a particle-hole transformation. 
Neglecting contributions from other states, we can approximate the sample Green function by $g_{M}(\omega,x,x)=\Phi_{0}(x)\Phi_{0}^{\dagger}(x)/\omega$. Note that we suppress position arguments throughout this section. Including the coupling to the tip the full Green function can be written as
\begin{equation}
G=\frac{1}{\omega-\tilde{\Sigma}_{M}}\Phi_{0}\Phi_{0}^{\dagger},
\end{equation}
where $\tilde{\Sigma}_{M}=\Phi_{0}^{\dagger}\Sigma\Phi_{0}$ is the self energy projected onto the Majorana bound state. We obtain
\begin{equation}
G^r_{eh}(\omega)=\frac{1}{\omega+i\pi t^{2}\abs{\zeta}^{2}\left[\rho(\omega_{+})+\rho(\omega_{-})\right]}\left(\begin{array}{cc}
u_{\uparrow}u_{\downarrow}^{*} & -u_{\uparrow}^{2}\\
u_{\downarrow}^{*2} & -u_{\downarrow}^{*}u_{\uparrow}.
\end{array}\right)
\end{equation}
where we introduced $\abs{\zeta}^{2}=\abs{u_{\uparrow}}^2 + \abs{u_{\downarrow}}^2$. Thus we find
\begin{equation}
\|G_{eh}(\omega)\|^{2}=\frac{\abs{\zeta}^{4}}{\omega^{2}+\pi^{2}t^{4}\abs{\zeta}^{4}\left[\rho(\omega_{+})+\rho(\omega_{-})\right]^{2}}
\end{equation}
and using Eq.~(\ref{eq:current}) we obtain the current 
\begin{equation}
I_{M}(V)=\frac{e}{h}\int d\omega\,\frac{\Gamma^M_{e}(\omega)\Gamma^M_{h}(\omega)}{\omega^{2}+\left(\Gamma^M_{e}(\omega)+\Gamma^M_h(\omega)\right)^{2}/4}\left[n_{F}(\omega-eV)-n_{F}(\omega+eV)\right],\label{eq:Majorana_current}
\end{equation}
with $\Gamma_{e/h}^{M}(\omega)=2\pi t^{2}\abs{\zeta}^{2}\rho(\omega_{\mp})$.

We now evaluate the current near the threshold $eV=\Delta+\eta$, $\eta\ll\Delta.$ The current at the opposite threshold $eV=-\Delta$ follows from $I(-V)=-I(V)$. 
At low temperatures, $T\ll\Delta$, we can set $n_{F}(\omega-eV)-n_{F}(\omega+eV)\sim 1$ and the Majorana current reads
\begin{equation} I_{M}(V)\simeq\frac{2e}{h}\int_{0}^{\eta}d\omega\,\frac{\Gamma_{e}^{M}(\omega)\Gamma_{h}^{M}(\omega)}{\omega^{2}+(\Gamma^{M}_{e}(\omega)+\Gamma^M_h(\omega))^{2}/4}.
    \label{eq:I_M}
\end{equation}
To lowest order in $\eta$, one can approximate
\begin{equation}
\rho(\omega\pm eV)\simeq\nu_{0}\sqrt{\frac{\Delta}{2}}\frac{\theta(\eta\pm\omega)}{\sqrt{\eta\pm\omega}}.
\label{eq:apprx_rho}
\end{equation}
This yields
\begin{equation}
    I_{M}(V)\simeq\frac{2e}{h}\eta (4-\pi)\Lambda(\eta/\omega_t)
    \label{eq:I_M2}
\end{equation}
as given in Eq.~(6) of the main text, where we have defined 
\begin{align}
\Lambda(x) =\frac{4}{4-\pi}\int_{0}^{1}dz\,\frac{1}{\sqrt{1-z^2}}\frac{1}{ z^2x^3+\left(\frac{1}{\sqrt{1+z}}+\frac{1}{\sqrt{1-z}}\right)^{2}}.
\end{align}
At the threshold we find
\begin{align}
\Lambda(0) =\frac{1}{4-\pi}\int_{0}^{1}dz\,\frac{2\sqrt{1-z^2}}{1+\sqrt{1-z^{2}}}=1
\end{align}
which yields Eq.~(1) of the main text. At large voltages, $\eta\gg \omega_t$, we instead find a negative differential conductance $dI/dV\propto -1/\eta^3$ in agreement with the lineshape shown in Fig.~(1) of the main text.

\subsection{Non-topological Andreev states at zero-energy}

A non-topological zero-energy Andreev bound state is characterized by two Nambu spinors 
\begin{equation}
    \Phi_{+}=\left(\begin{array}{c}
        \mathbf{u}\\
        \mathbf{v}
    \end{array}\right),\quad\Phi_{-}=\left(\begin{array}{c}
         \Theta\mathbf{v}\\
        -\Theta \mathbf{u}
    \end{array}\right)
\end{equation}
where $\mathbf{u}=(u_{\uparrow},u_{\downarrow})^{\rm T}$,  $\mathbf{v}=(v_{\downarrow},v_{\uparrow})^{\rm T}$ are functions of space and $ \Theta=-i\sigma_{y}{\cal K}$ is the time-reversal operator with $\cal{K}$ the complex conjugation.
The Lehmann representation of the real space Green function is thus a  $4\times 4$ Matrix in Nambu and spin space
\begin{equation}
    g(\omega)=\frac{\Phi_{+}\Phi_{+}^{\dagger}+\Phi_{-}\Phi_{-}^{\dagger}}{\omega}=\frac{1}{\omega}(\begin{array}{c}
        \Phi_{+} , \Phi_{-}\end{array})\left(\begin{array}{c}
        \Phi_{+}^{\dagger}\\
        \Phi_{-}^{\dagger}
    \end{array}\right).
    \label{eq:2ABS_GF}
\end{equation}
When spin is a good quantum number the Green function may be reduced to a $2\times2$ Matrix in particle-hole space only. In the case of pure $s$-wave pairing, where Cooper pairs are formed from electrons with opposite spin, we can set $u_{\downarrow}=v_{\uparrow}=0$. The spinors $\Phi_{+}$ and $\Phi_-$ then belong to the orthogonal subspaces spanned by $(c_{\uparrow},0,c_{\downarrow}^\dagger,0)$ and $(0,c_{\downarrow},0,-c_{\uparrow}^\dagger)$. The Green function decomposes into two $2\times 2$ blocks, which are related by particle-hole symmetry and have equal contributions to the current. 
In the opposite case of a spin polarized $p$-wave superconductor we can set $u_{\downarrow}(x)=v_{\downarrow}(x)=0$. Now $\Phi_{+}$ and $\Phi_-$ belong to the same subspace spanned by $(c_{\uparrow},0,0,-c_{\uparrow}^\dagger)$ and the Green function is a single $2\times 2$ matrix.

It is therefore useful to first discuss a general $2\times 2$ Nambu Green function
\begin{align}
 g=\begin{pmatrix}g_{ee}&g_{eh}\\ g_{he} &g_{hh}\end{pmatrix}.
\end{align}
We will return to the $4\times 4$ case in Sec.~\ref{sec:spin-orbit} when discussing superconductors with spin-orbit coupling. The coupling to the tip can be included through the self-energy $\Sigma^r(\omega)= -i \pi t^2 \diag (\rho(\omega_-),\rho(\omega_+))$. The Green function of the coupled system is obtained from the Dyson equation $G=g(1-\Sigma g)^{-1}$.
We find
\begin{equation}
    G_{eh}^r(\omega)=\frac{g_{eh}(\omega)\omega}{\omega-\lambda(\omega)+i\pi\omega t^2(g_{ee}(\omega)\rho(\omega_-)+g_{hh}(\omega)\rho(\omega_+))}\label{green_func_2by2}
\end{equation}
where $\lambda(\omega)=\omega\pi^2 t^4\rho(\omega_+)\rho(\omega_-)\det g(\omega)$.
Using Eq.~(\ref{eq:current}), we obtain the current
\begin{align}
    I(V)=\frac{4\pi^2 e t^{4}}{h}\int_{\Delta-eV}^{eV-\Delta} d\omega\,
\frac{\omega^2\abs{g_{eh}(\omega)}^2\rho(\omega_-)\rho(\omega_+)}{[\omega-\lambda(\omega)]^2+\pi^2\omega^2 t^4\left[(g_{ee}(\omega)\rho(\omega_-)+g_{hh}(\omega)\rho(\omega_+)\right]^2}
\label{eq:current_2by2}.
\end{align}

We now calculate the conductance for a zero-energy Andreev state in the limiting cases of pure $s$-wave and spinless $p$-wave pairing and for a general spin-structure of the order parameter in the presence of spin-orbit coupling. The result for the three cases are compared in Fig.~\ref{fig:generic_conductance}.

\subsubsection{\texorpdfstring{\protect{$s$-wave pairing}}{}} \label{sec:s-wave-pairing}

When $u_\downarrow=v_\uparrow=0$ the Green function in Eq.~(\ref{eq:2ABS_GF}) is block diagonal.
The $2\times2$ block in the basis $(c_{\uparrow},c_{\downarrow}^\dagger)$ reads
\begin{equation}
    g=\frac{1}{\omega}
    \left(\begin{array}{cc}
        \abs{u_{\uparrow}}^{2} & u_{\uparrow}v_{\downarrow}^{*}\\
        u_{\downarrow}^{*}v_{\uparrow} & \abs{v_{\downarrow}}^{2}
    \end{array}\right).
\end{equation}
We find $\det g(\omega)=0$ and thus $\lambda(\omega)$ vanishes. Using Eq.~(\ref{eq:current_2by2}), we arrive at the current as given in Eq.\ (9) of the main text, 
\begin{equation}
   I_A(V)=\frac{2e}{h}\int_{\Delta-eV}^{eV-\Delta} d\omega\,
\frac{\Gamma_e^A(\omega)\Gamma_h^A(\omega)}{\omega^2+\left(\Gamma_e^A(\omega)+\Gamma_h^A(\omega)\right)^2/4}
\end{equation}
where $\Gamma_e^A(\omega)=2\pi t^2 \abs{u_\uparrow}^2\rho(\omega_-)$, $\Gamma_h^A(\omega)=2\pi t^2\abs{v_{\downarrow}}^2\rho(\omega_+)$. We have included an extra factor of two to account for the second $2\times2$ block of the Green function which yields an equal contribution as a consequence of particle-hole symmetry. Near the threshold when $eV=\Delta+\eta$, $\eta\ll\Delta$ we find
\begin{align}
    I_A(V)=2I_M(V)f(|u(r)|^2/|v(r)|^2), 
\end{align}
which can be obtained by a similar analysis as for the Majorana bound state in the previous section. The dimensionless function 
\begin{equation}
  f(x)=\frac{2x}{4-\pi}\int_{-1}^1 dz \sqrt{1-z^2}/(x\sqrt{1-z}+\sqrt{1+z})^2  
\end{equation}
takes on values between 0 and 1. Thus, the threshold differential conductance is
\begin{equation}
\left.\frac{dI_{A}}{dV}\right|_{eV=\Delta}=\frac{4e^{2}}{h}f(|u(r)|^2/|v(r)|^2)(4-\pi).
\end{equation}

\subsubsection{\texorpdfstring{\protect{$p$-wave pairing}}{}}\label{sec:p-wave-pairing}

In a spin-polarized $p$-wave superconductor Cooper pairs we can set $u_\downarrow=v_\downarrow=0$.
The Green function of an Andreev state reduces to a single $2\times2$ matrix 
\begin{equation}
    g=\frac{1}{\omega}\left(\begin{array}{cc}
        \abs{u_{\uparrow}}^{2}+\abs{v_{\uparrow}}^{2} & 2u_{\uparrow}v_{\uparrow}^{*}\\
        2u_{\uparrow}^{*}v_{\uparrow} & \abs{u_{\uparrow}}^{2}+\abs{v_{\uparrow}}^{2}
    \end{array}\right).
\end{equation}
We have introduced $\lambda(\omega)=\pi^2t^4\rho(\omega_+)\rho(\omega_-)(\abs{u_\uparrow}^2-\abs{v_\uparrow}^2)^2/\omega$ and using Eq.~(\ref{eq:current_2by2}) we obtain
\begin{equation}
I_A(V)=\frac{ 16 \pi^2 et^{4}}{h}\int_{-\eta}^{\eta} d\omega\,
    \frac{\abs{u_{\uparrow}v_{\uparrow}}^{2}\rho(\omega_{-})\rho(\omega_{+})}{\left[\omega-\frac{\rho(\omega_{-})\rho(\omega_{+})\pi^{2}\abs{t^{4}}}{\omega}\left(\abs{u_{\uparrow}}^{2}-\abs{v_{\uparrow}}^{2}\right)^2\right]^{2}+\pi^{2}t^{4}(\abs{u_{\uparrow}}^{2}+\abs{v_{\uparrow}}^{2})^{2}[\rho(\omega_{+})+\rho(\omega_{-})]^{2}}
\end{equation}
with $eV=\Delta+\eta$. Close to the threshold $\eta \to 0$ the denominator is dominated by $\lambda(\omega)\propto 1/\eta^2$ and we generically find 
$I\propto \eta^4\theta(\eta)$. Hence the conductance $G\propto \eta^3\theta(\eta)$ is continuous in contrast to the case of $s$-wave pairing. 
At certain points in space it may be possible that $|u_\uparrow|=|v_\uparrow|$ in which case the conductance still jumps at the threshold. Slightly moving away from such points should restore the smooth onset of the conductance at the threshold. 
The conductance exhibits a peak at the characteristic scale $\eta\sim\omega_t$ with $\omega_t=(\max\{|u_\uparrow|^2,|v_\uparrow|^2\}\nu_0t^2\sqrt{\Delta})^{2/3}$ as shown in Fig.~\ref{fig:generic_conductance}.

\subsubsection{Generic case with spin-orbit coupling}\label{sec:spin-orbit}

Realistic proposals of topological superconductors typically involve a mixture of $s$-wave and $p$-wave pairing. In particular, such pairing arises in any superconductor with spin-orbit coupling. In this case all components of $\bf{u}$ and $\bf{v}$ are generically nonzero. The full sample Green function including the coupling to the tip due to the self-energy in Eq.~(\ref{eq:self-energy}) can be written in terms of Dyson series
\begin{align}
    G &=\frac{1}{\omega}\left(\begin{array}{cc}
        \Phi_{+} & \Phi_{-}\end{array}\right)\left[1+\frac{1}{\omega}\left(\begin{array}{c}
            \Phi_{+}^{\dagger}\\
            \Phi_{-}^{\dagger}
        \end{array}\right)\Sigma\left(\begin{array}{cc}
        \Phi_{+} & \Phi_{-}\end{array}\right)+\dots\right]\left(\begin{array}{c}
        \Phi_{+}^{\dagger}\\
        \Phi_{-}^{\dagger}
    \end{array}\right) .
\end{align}
A straightforward calculation reveals
\begin{equation}
    G^{r}_{eh}=\frac{\left[(\omega-\tilde{\Sigma}_{--}^{r})\mathbf{u}+\tilde{\Sigma}_{-+}^{r}({\cal
    C}\mathbf{v})\right]\mathbf{v}^{\dagger}-\left[(\omega-\tilde{\Sigma}_{++}^{r}){\cal C}\mathbf{v}+\tilde{\Sigma}_{+-}^{r}\mathbf{u}\right]({\cal
    C}\mathbf{u})^{\dagger}}{\omega^{2}-\omega(\tilde{\Sigma}_{++}^{r}+\tilde{\Sigma}_{--}^{r})+\tilde{\Sigma}_{++}^{r}\tilde{\Sigma}_{--}^{r}-\tilde{\Sigma}_{+-}^{r}\tilde{\Sigma}_{-+}^{r}}
\end{equation}
where the projected self-energies are 
\begin{subequations}
\begin{gather}
    \tilde{\Sigma}^r_{++}(\omega)=-i \pi t^{2}\left[\|\mathbf{u}\|^{2}\rho(\omega_{-})+\|\mathbf{v}\|^{2}\rho(\omega_{+})\right]  \\
    \tilde{\Sigma}^r_{--}(\omega)=-i \pi t^{2}\left[\|\mathbf{v}\|^{2}\rho(\omega_{-})+\|\mathbf{u}\|^{2}\rho(\omega_{+})\right] \\
    \tilde{\Sigma}^r_{+-}(\omega)=-i \pi t^{2}\langle\mathbf{u},\Theta\mathbf{v}\rangle\left[\rho(\omega_{-})+\rho(\omega_{+})\right]  \\
    \tilde{\Sigma}^r_{-+}(\omega)=-i \pi t^{2}\langle \Theta\mathbf{v},\mathbf{u}\rangle\left[\rho(\omega_{-})+\rho(\omega_{+})\right]=-\tilde{\Sigma}^r_{+-}(\omega)^{*},
\end{gather}
\end{subequations}
and $\langle \cdot , \cdot \rangle$ is the inner product. In the above derivations, we have used the anti-unitarity of the time-reversal operator
and that $\Theta^2=-1$, namely $\langle {\bf u},\Theta{\bf v}\rangle=\langle {\bf v},\Theta^\dagger{\bf u}\rangle = -\langle {\bf v},\Theta{\bf u}\rangle$.
Then the norm can be written as
\begin{equation}
    \|G_{eh}\|^{2}=W\frac{2\omega^{2}+W\pi^{2}t^{4}Y\left\{[\rho(\omega_{+})^{2}+\rho(\omega_{-})^{2}]Z+4\rho(\omega_{+})\rho(\omega_{-})Y\right\}}{\left\{
        \omega^{2}-W\pi^{2}t^{4}\left[(\rho(\omega_{+})^{2}+\rho(\omega_{-})^{2})Y+Z\rho(\omega_{+})\rho(\omega_{-})\right]\right\}
        ^{2}+W\omega^{2}\pi^{2}t^{4}(Z+2)[\rho(\omega_{-})+\rho(\omega_{+})]^{2}}
\end{equation}
where 
\begin{subequations}
    \begin{gather}
        W=\|\mathbf{u}\|^{2}\|\mathbf{v}\|^{2}+\abs{\langle\mathbf{u},\Theta\mathbf{v}\rangle}^{2}\\
        Y=\left(\|\mathbf{u}\|^{2}\|\mathbf{v}\|^{2}-\abs{\langle\mathbf{u},\Theta \mathbf{v}\rangle}^{2}\right)/W\\
        Z=\left(\|\mathbf{u}\|^{4}+\|\mathbf{v}\|^{4}-2\abs{\langle\mathbf{u},\Theta \mathbf{v}\rangle}^{2}\right)/W.
    \end{gather}
\end{subequations}
The differential conductance is then a function of $Y$, $Z$ and $\eta/\tilde{\omega}_t$, where $\tilde{\omega}_t^3 =\Delta\nu_{0}^{2}\pi^{2}t^{4}W$.
The parameter $Y$ interpolates between $s$-wave pairing ($Y=1$, Sec.~\ref{sec:s-wave-pairing}), where the threshold conductance is maximal, $p$-wave pairing ($Y=0$, Sec.~\ref{sec:p-wave-pairing}), where the threshold conductance is zero. 
\begin{figure}[t]
    \includegraphics[width=0.6\textwidth]{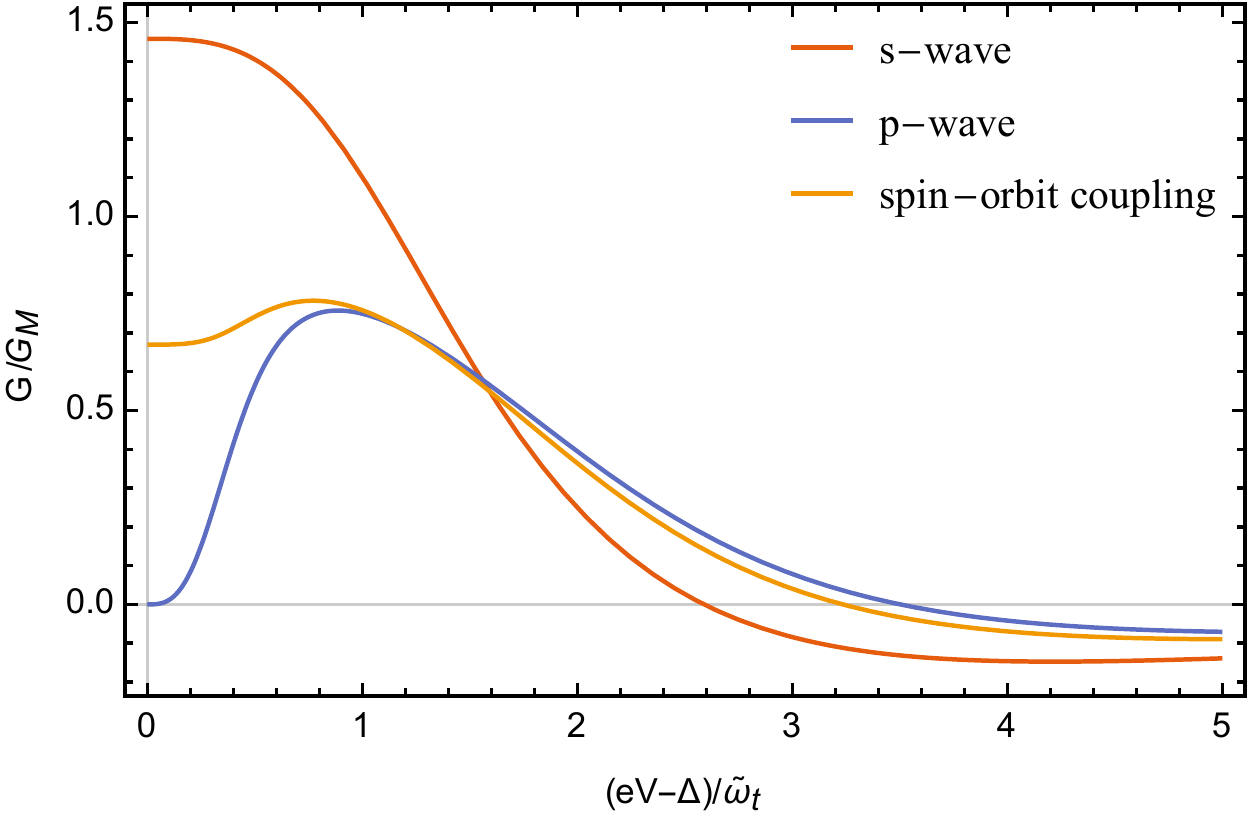}
    \caption{Conductance near the threshold. The parameters for the $s$-wave pairing case are $Y=1$, $Z=4.25$.
    For the $p$-wave pairing case, $Y=0$, $Z=1.13$. And in the case with spin-orbit coupling, $Y=0.05$, $Z=0.74$.}
    \label{fig:generic_conductance}
\end{figure}
In Fig.~\ref{fig:generic_conductance} we show the conductance for different values of this parameter. 

\section{Continuum contribution to the conductance}

In this section, we provide details on evaluating the contribution of the quasiparticle continuum to the conductance near the threshold. We identify two different effects of the continuum states, namely, (a) a possible smoothening of the step in the bound-state conductance [see Fig.~1(a) and Eq.~(10) of the main text] due to interference between resonant and nonresonant Andreev reflections and (b) additonal contributions to the conductance from nonresonant Andreev reflections. The latter effect becomes important when probing the conductance away from the bound state, where resonant and nonresonant Andreev reflections compete as discussed in the main text.

\subsection{Continuum effect on resonant Andreev reflections}
\subsubsection{General considerations}

As follows from Sec.~\ref{sec:bound_state_conductance}, the lineshape of a zero-energy Andreev state in an $s$-wave superconductor closely resembles that of a Majorana state. In particular, the conductance exhibits a step at the threshold $eV=\Delta$ in both cases. We now revisit these two cases and discuss whether this step is robust when the quasiparticle continuum is taken into account. As discussed in the main text, we find a suppression of the conductance for the Andreev state close to the threshold, while the step remains robust for the Majorana. This behavior is reminiscent of tunneling from a normal metal tip, where Majorana states appear as a robust zero-bias conductance peaks while zero-energy Andreev states generically exhibit zero conductivity.

For simplicity, we focus our analysis on samples in which spin is a good quantum number and that can be described in $2\times 2$ Nambu space. An example of a more general model is discussed in Sec.~\ref{sec:QSHI_edge}. In particular, we consider an $s$-wave superconductor with a zero-energy Andreev state and a spinless $p$-wave superconductor with a Majorana state. In these cases the Lehmann representation of the Green function reads 
\begin{equation}
    g(\omega) = \frac{\Phi_0(x)\Phi_0^{\dagger}(x)}{\omega} + \sum_{n} \int_{\abs{E}\geq\Delta_s} dE  \frac{\Phi_{E,n}(x)\Phi^\dagger_{E,n}(x)}{\omega-E},
\end{equation}
where the spinor $\Phi_{0}$ describes a single zero-energy bound state and $\Phi_{E,n}$ are continuum states above the gap $\Delta_s$ with energy $E$ and index $n$ labeling the degeneracy. 
Using this expression in the Green function in the presence of a tunnel coupling in Eq.~(\ref{green_func_2by2}) and expanding to lowest order in energy, we obtain
\begin{equation}
    G_{eh}^r(\omega)=\frac{\Phi_0(r) \Phi_0^\dag(r)}{\omega-\lambda(\omega)+i(\Gamma_e+\Gamma_h)/2}.
\end{equation}
To determine the effect of the extra term $\lambda(\omega)\propto \det g(\omega)$ we analyze the energy dependence of $\det g(\omega)$ in more detail. The function $g(\omega)$ has a simple pole at $\omega =0$ and branch cuts on the real axis for $|\omega|>\Delta_s$ but is analytic elsewhere. 
The residue of $g(\omega)$ at $\omega=0$ is $\Phi_0(x)\Phi_0^\dag(x)$ which has zero determinant and we can thus write
\begin{equation}
    \lim_{\omega \to 0} \omega^2 \det g(\omega) = 0
\end{equation}
and expand $\det g(\omega)$ as a Laurent series around $\omega=0$ 
\begin{equation}
    \det g(\omega) = \sum_{n=-1}^{\infty}c_{n}\omega^n.
    \label{eq:laurent}
\end{equation}
Furthermore $g(\omega)$ is hermitian at nonzero subgap energies and thus its determinant is real. We now determine the lowest-order cotribution to $\det g$ for Majorana and Andreev states.

\subsubsection{Majorana states}\label{sec:continuum_Majorana}

The Green function of a topological superconductor satisfies particle-hole symmetry $U_{\cal C}g(\omega)U^\dag_{\cal C}=-g^*(-\omega)$ with a unitary operator $U_{\cal C}$. This yields $\det g(\omega)=-\det g^*(-\omega)$ and reality further requires $\det g(\omega)$ to be an even function of $\omega$. Hence, we obtain $c_{-1}=0$ and $\lambda(\omega)\sim \omega t^4\rho(\omega_+)\rho(\omega_-)$, which approaches a constant in the limit $\eta=eV-\Delta\to 0^+$ with $|\omega|<\eta$. The Majorana contribution to the current reads
\begin{equation}
    I_{M}(V)\simeq\frac{2e}{h}\int_{0}^{\eta}d\omega\,\frac{\Gamma_{e}^{M}(\omega)\Gamma_{h}^{M}(\omega)}{(\omega-\lambda)^{2}+(\Gamma^{M}_{e}(\omega)+\Gamma^M_h(\omega))^{2}/4}.
\end{equation}
The rates $\Gamma^M_{e/h}$ diverge at the threshold and the continuum term $\lambda(\omega)$ becomes negligible. The threshold conductance of Majorana states thus remains unaffected by the continuum states.

\subsubsection{Zero-energy Shiba state}

In contrast to Majoranas, the $2\times2$ Green function describing zero-energy Andreev states generically does not satisfy particle-hole symmetry and thus $c_{-1}\neq 0$. As an example of a system with trivial zero-energy Andreev state, we consider a magnetic impurity in an $s$-wave superconductor, which induces a bound state localized at the impurity, known as a Shiba state. The Hamiltonian describing a Shiba state localized at the origin due to a magnetic impurity can be written in first quantization as
\begin{equation}
    \mathcal{H}_S(\v{x}) = \mathcal{H}_{BCS}(\v{x}) + (V\tau_z-JS\sigma_z)\delta(\v{x}),
\end{equation}
where $\tau_i$ and $\sigma_i$ are Pauli matrices in Nambu and spin space, where $V$ and $JS$ are the potential scattering and exchange coupling strength. Since ${\cal H}_S$ is block diagonal in spin space, we only need to deal with one of the blocks, say $\sigma_z=1$. The other block related to the first by particle-hole symmetry contributes a second (equivalent) zero-energy state as mentioned above. The sample Green function $g_S$ at the impurity position has the form \cite{Ruby2015a}
\begin{align}
    g_S(\omega)
    &= \frac{\pi\nu_s}{2\omega\alpha - (1-\alpha^2+\beta^2)\sqrt{\Delta_s^2-\omega^2}}
    \left(\begin{array}{cc}
        \omega+(\alpha+\beta)\sqrt{\Delta_s^{2}-\omega^{2}} & \Delta_s\\
        \Delta_s & \omega+(\alpha-\beta)\sqrt{\Delta_s^{2}-\omega^{2}}
    \end{array}\right), 
\end{align}
where $\nu_s$ and $\Delta_s$ are the normal density of states and gap of the sample and we introduced the dimensionless parameters $\alpha=\pi\nu_{0}JS$ and $\beta= \pi \nu_0 V$. 
The Shiba bound state energy is given by 
\begin{align}
      \epsilon_{0}=\Delta_s\frac{1-\alpha^{2}+\beta^2}{\sqrt{(1-\alpha^2+\beta^2)^2+4\alpha^2}}.
\end{align}
Thus, since we focus on zero-energy bound states we set $\alpha^2=1+\beta^2$. After a straightforward calculation, we obtain
\begin{equation}
    \omega \det g(\omega) \simeq \frac{\pi^2\nu_s^2\Delta_{s}}{2\alpha}
\end{equation}
for small $\omega$. From the Green function we also obtain (for $V>0$)
\begin{gather}
    \abs{u}^2 = \pi \nu_s \Delta_s (\alpha +\sqrt{\alpha^2-1}),\\
    \abs{v}^2 = \pi \nu_s \Delta_s (\alpha -\sqrt{\alpha^2-1}).
\end{gather}
Near the threshold, for small  $\eta=eV-\Delta$, the term $\propto \lambda(\omega)$ in the denominator of Eq.~(\ref{eq:current_2by2}) dominates and we find
\begin{equation}
    I_A(V) = \frac{2e}{h}\frac{16\eta^{2}\alpha^{2}}{\nu_{0}^{2}\nu_{s}^{2}\pi^{3}t^4\Delta}.
\end{equation}
Thus the conductance $dI_A/d\eta$ is zero at $eV=\Delta$ and then rises linearly as described by Eq.~(10) of the main text.
This originates from the interference between resonant and nonresonant Andreev reflection, which suppresses the differential conductance at $eV=\Delta$. 
In order to estimate the width of this suppression we determine the value of $\eta$ when the term $\lambda$ in the denominator of Eq.~(\ref{eq:current_2by2}) becomes of the same order of magnitude as the tunneling broadening $\sim\omega t^2[ g_{ee}\rho(\omega_-)+ g_{hh}\rho(\omega_+)]$, i.e.,
\begin{equation}
    \frac{\nu_{0}^{2}\Delta t^4\omega\det g(\omega)}{\eta}  \sim \nu_0|t|^2\left(\abs{u}^2 +\abs{v}^2\right)\sqrt{\frac{\Delta}{\eta}},
\end{equation}
which yields
\begin{equation}
    \eta \sim \frac{\nu_0^2\nu_s^2t^4\Delta}{\alpha^4}\sim \frac{{\cal T}^2\Delta}{\alpha^4}.
\end{equation}

\subsection{Nonresonant Andreev reflection}

We now determine the conductance due to nonresonant Andreev reflections from the quasiparticle continuum in the substrate. 
To evaluate the conductance contributions from the continuum, we focus on a specific model of a Dirac Hamiltonian with a domain wall, where the mass changes sign. We first calculate the real space Green function and then evaluate the conductance as a function of separation from the impurity.

\subsubsection{Topological superconductor Green function} \label{sec:Majorana_Dirac}

As an example of a topological superconductor hosting a Majorana state we calculate the Green function of a Dirac Hamiltonian with a domain wall
 \begin{equation}
     H_{\rm Dirac}=-iv_F\partial_x \tau_x - m[\theta(x)-\theta(-x)]\tau_z\label{Dirac_Hamiltonian}
 \end{equation}
in the spinless Nambu basis $(\psi,\psi^\dag)$. Here $v_F$ is the velocity of the Dirac fermion and $m$ is the effective mass. The system satisfies particle-hole symmetry $\{\mathcal{C},H_{\rm Dirac}\}=0$ with the charge conjugation operator $\mathcal{C}=\tau_x \mathcal{K}$, where $\mathcal{K}$ is the complex conjugation. This is a generic model for the low-energy behavior of a topological superconductor close to the phase transition. Specifically, a spinless $p$-wave superconductor with a chemical potential close to the bottom of the band can be approximated by Eq.~(\ref{Dirac_Hamiltonian}). To compute the eigenstates of this Hamiltonian it is convenient to solve for the rotated Hamiltonian $e^{i\pi\tau_y/4}H_{\rm Dirac}e^{-i\pi\tau_y/4}$. Notice that the charge conjugation operator is now given by $\tau_z \mathcal{K}$.

One readily shows that the domain wall supports a single Majorana state at zero energy with the wavefunction 
\begin{equation}
    \psi_{0}(x)=\sqrt{\frac{m}{2v_F}}e^{-m|x|/v_F}
    \left(\begin{array}{c}
        i\\
        1
    \end{array}\right).
\end{equation}
All other states in the system are extended with energies above the gap $m$. The wavefunctions of the continuum states can be determined using the ansatz 
 \begin{align}
     \psi_{\pm\epsilon}(x) &= 
     \left( \begin{array}{c}
         m{\rm sgn}\{x\} \\ -(v_Fp\mp \epsilon) \end{array}
     \right)e^{ipx}
     \left[A\theta(-x)+B\theta(x)
     \right]
     +      \left( \begin{array}{c}
         m{\rm sgn}\{x\} \\ (v_Fp\pm \epsilon) \end{array}
     \right)e^{-ipx}
     \left[C\theta(-x)+D\theta(x)
     \right],
 \end{align}
where $v_Fp = \sqrt{\epsilon^2-m^2}$ and $\epsilon$ is the positive eigenvalue. 
The matching condition at $x=0$ leads to 
 \begin{equation}
     A+B+C+D=0,\qquad  v_Fp(A-B-C+D)\mp\epsilon(A-B+C-D)= 0.
 \end{equation}
For positive-energy solutions, we get
 \begin{equation}
     \frac{A+D}{A+C} = \frac{\epsilon}{v_Fp} = \sqrt{1+\left(\frac{m}{v_Fp}\right)^2}.
 \end{equation}
For incoming waves from $x=-\infty$ we require $D=0$ and obtain
 \begin{equation}
     C^{in} = \frac{v_Fp-\epsilon}{\epsilon}A^{in}, \qquad
     B^{in} = -\frac{v_Fp}{\epsilon}A^{in}.
 \end{equation}
Similarly, for outgoing waves to $x=-\infty$ we have $A=0$ and thus 
 \begin{equation}
     C^{out} = \frac{v_Fp}{\epsilon}D^{out}, \qquad  B^{out} = -\frac{\epsilon+v_Fp}{\epsilon}D^{out}.
 \end{equation}
This yields the scattering states
\begin{align}
        \psi^{in}_\epsilon(x) &=
        A
        \left\{
        \left( \begin{array}{c}
                        -m\theta(-x) - m \frac{v_Fp}{\epsilon}\theta(x)
                        \\
                        (\epsilon-v_Fp)\theta(-x)-\frac{v_Fp}{\epsilon}(\epsilon-v_Fp)
                        \theta(x)
                        \end{array} 
        \right)e^{ipx} + \frac{v_Fp-\epsilon}{\epsilon}
        \left( \begin{array}{c}
                        -m
                        \\
                        v_Fp+\epsilon
                        \end{array}
        \right)\theta(-x)e^{-ipx}
        \right\}, \nonumber \\
                \psi^{out}_\epsilon(x) &=
                D
        \left\{\frac{\epsilon+v_Fp}{\epsilon}
        \left( \begin{array}{c}
                        -m
                        \\
                        v_Fp-\epsilon
                        \end{array}
        \right)\theta(x)e^{ipx} +
        \left( \begin{array}{c}
                        m\theta(x)-m\frac{v_Fp}{\epsilon}\theta(-x)
                        \\
                        (v_Fp+\epsilon)\theta(x)+v_Fp\frac{v_Fp+\epsilon}{\epsilon}
                        \theta(-x)
                        \end{array}
        \right)e^{-ipx}
        \right\}.        
        \label{eq.in_out_state}
\end{align}
One can easily check that incoming and outgoing states are orthogonal $ \langle \psi^{out}_\epsilon | \psi^{in}_\epsilon \rangle=0$. 
From the normalization conditions $\|\psi_{\epsilon}^{in}\|=\|\psi_{\epsilon}^{out}\|=1$ we obtain the coefficients $A = [2L\epsilon(\epsilon-v_Fp)]^{-1/2}$ and $D = [2L\epsilon(\epsilon+v_Fp)]^{-1/2}$ for a system of size $L$. The Green function can now be obtained from its spectral decomposition
\begin{equation}
        \tilde{g}(x,x';\omega) =
        \frac{\psi_0(x)\psi_0(x')^\dagger}{\omega}
         + 
        \sum_{\substack{\epsilon>0,\\ \alpha=in,out}}\bigg[
        \frac{\psi^{\alpha}_\epsilon(x)\psi_\epsilon^{\alpha}(x')^\dagger}{\omega-\epsilon}
        +
        \frac{\left(\mathcal{C}\psi^{\alpha}_\epsilon(x)\right)\left(\mathcal{C}\psi_\epsilon^{\alpha}(x')\right)^\dagger}{\omega+\epsilon}
    \bigg] \label{eq:GF_rotated}
\end{equation}
where we have used the charge conjugation operator $\mathcal{C}$ to relate the negative solution to the positive ones. Focusing on $x=x'=0$ for simplicity we find the wavefunctions of the continuum states
\begin{equation}
    \psi^{in}_\epsilon(0) =
    -\frac{1}{\sqrt{2L\epsilon(\epsilon-v_Fp)}}
    \frac{v_Fp}{\epsilon}
    \left(
        \begin{array}{c}
            m \\
            \epsilon-v_Fp
        \end{array}
    \right),\qquad
    \psi^{out}_\epsilon(0) =
    \frac{1}{\sqrt{2L\epsilon(\epsilon+v_Fp)}}
    \frac{v_Fp}{\epsilon}
    \left(
        \begin{array}{c}
            -m \\
            \epsilon+v_Fp
        \end{array}
    \right).
\end{equation}
Using Eq.~(\ref{eq:GF_rotated}), we obtain the Green function
\begin{equation}
    \tilde{g}(0,0;\omega)=
    \frac{m}{2v_F\omega}\left(\begin{array}{cc}
        1 & i\\
        -i & 1
    \end{array}\right)
    +
    \frac{1}{L}
    \sum_{\epsilon>0}
    \frac{\epsilon^2-m^2}
    {\epsilon^2}
    \frac{2\omega}{\omega^2-\epsilon^2}
    \left(
    \begin{array}{cc}
        1 & 0
        \\
        0 & 1
    \end{array}
    \right)
    = \frac{1}{2v_F}\frac{m}{\omega}
    \left(
    \begin{array}{cc}
        \sqrt{1-(\omega/m)^2} & i \\
        -i & \sqrt{1-(\omega/m)^2}
    \end{array}
    \right).
\end{equation}
Here we have used 
\begin{equation}
    \frac{1}{L}\sum_{\epsilon>0} = 
    \int\!\frac{dp}{2\pi} = \frac{1}{2\pi v_F} \int_{m}^{\infty}\!d\epsilon
    \frac{\epsilon}{\sqrt{\epsilon^2-m^2}}.
\end{equation}
Notice that the matrix commutes with $\tau_y$, and thus commutes with the rotation operation introduced at the beginning.
We find the Green function for the original Hamiltonian $H_{\rm Dirac}$
\begin{equation}
    g(0,0;\omega) 
    = \pi\nu_s \frac{m}{\omega}
    \left(
    \begin{array}{cc}
        \sqrt{1-(\omega/m)^2} & i \\
        -i & \sqrt{1-(\omega/m)^2}
    \end{array}
    \right)
    \label{eq:GF_Dirac},
\end{equation}
where $\nu_s=(2\pi v_F)^{-1}$ is the normal density of states. 

Away from the mass domain wall, the sample Green function at $x\geq 0 $ can be computed similarly from Eq.~(\ref{eq.in_out_state}) and (\ref{eq:GF_rotated}), which yields
\begin{align}
    g_{ee}(x)&=g_{hh}(x)=\frac{\pi\nu_{s}}{\sqrt{1-(\omega/m)^{2}}}\left(\frac{m}{\omega}e^{-2 x/\xi}-\frac{\omega}{m}\right)
    \nonumber \\
    g_{eh}(x)&=g_{he}(x)^*=\pi\nu_{s}\left(i\frac{m}{\omega}e^{-2 x/\xi}-\frac{1-e^{-2 x/\xi}}{\sqrt{1-(\omega/m)^{2}}}\right).
    \label{eq.gf_position}
\end{align}
where $\xi=[2\pi\nu_{s}m\sqrt{1-(\omega/m)^{2}}]^{-1}=v_F/m\sqrt{1-(\omega/m)^{2}}$ is the coherence length that characterizes the exponential decay of the Majorana wavefunction. 

\subsubsection{Differential conductance away from the Majorana state}

\begin{figure}[t]
    \includegraphics[width=0.6\textwidth]{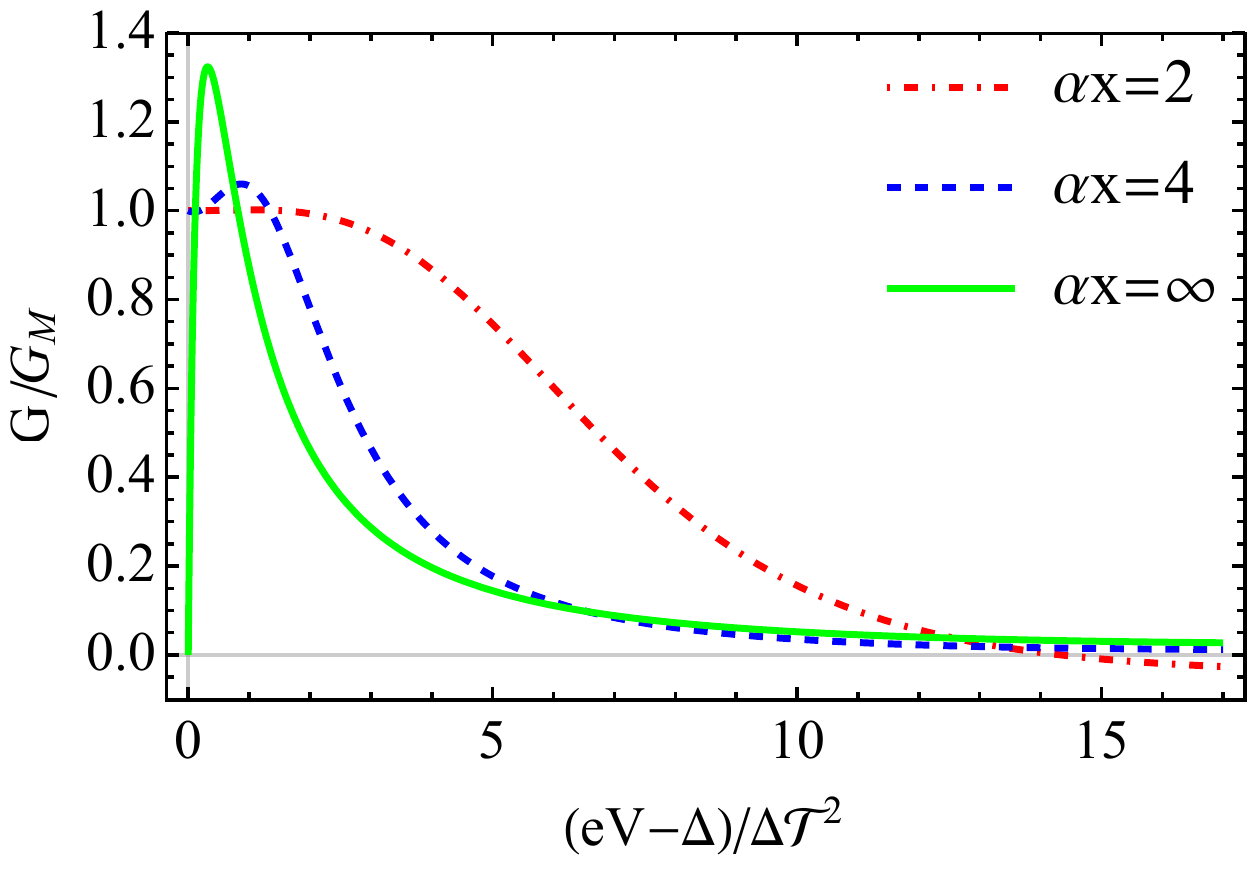}
    \caption{Differential conductance of the topological superconductor with a mass domain wall. 
        The three curves correspond different tip locations measured from the domain wall. For all curves we have set $\Delta\mathcal{T}^2/m=0.016$.}
    \label{fig.position_conductance}
\end{figure}

The Green function has a pole at zero energy whose amplitude decays exponentially away from the domain wall reflecting the probability density of the Majorana bound state. This singular part fully determines the threshold conductance and yields the quantized value $G_M$.
Notice that $\lim_{\omega\rightarrow 0}\omega\det g =0$ guarantees the robustness of this quantization consistent with the general argument based on particle-hole symmetry given in Sec.~\ref{sec:continuum_Majorana}. 
The conductance at voltages above the threshold depends on the competition between the singular part and a nonsingular contribution which originates from the quasiparticle continuum. The continuum contribution can be readily evaluated for $x\to \infty$, where the Green function describes a homogeneous $p$-wave superconductor,
\begin{equation}
    g(\omega)=-\frac{\pi\nu_{s}}{\sqrt{m^{2}-\omega^{2}}}\left(\begin{array}{cc}
        \omega & m\\
        m & \omega
    \end{array}\right).
\end{equation}
Using this expression in the current formula in Eq.~(\ref{eq:current_2by2}) we find 
\begin{align}
    I&=\frac{4\pi^{2}e\abs{t}^{4}}{h}\int^\eta_\eta
    d\omega\frac{\rho_{+}\rho_{-}\pi^{2}\nu_{s}^{2}m^{2}}{\left(m^{2}-\omega^{2}\right)(1+\pi^{4}\nu_{s}^{2}t^{4}\rho_{+}\rho_{-})^{2}+\pi^{4}\nu_{s}^{2}t^{4}\omega^{2}(\rho_{+}+\rho_{-})^{2}}\nonumber
    \\
    & \simeq  \frac{8e}{h}\Delta{\cal T}^{2}\int_{-1}^{1}dz\frac{\sqrt{1-z^{2}}}{\left(2\sqrt{1-z^{2}}+\Delta{\cal T}^2/\eta\right)^{2}},
\end{align}
where we introduced the transmission probability of the junction $\mathcal{T}=\pi^2\nu_{s}\nu_{0}t^2$ and assumed $\Delta{\cal T}^2\ll m$.
This yields the conductance
\begin{equation}
    G(V)=\frac{8G_{M}}{4-\pi}\int_{0}^{1}dz\,\frac{(2\eta/\Delta{\cal T}^2)\sqrt{1-z^{2}}}{(1+(2\eta/\Delta{\cal T}^2)\sqrt{1-z^{2}})^{3}}.
\end{equation}
The conductance has a maximum at $\eta\sim\Delta{\cal T}^2$ with a magnitude $\sim 1.3 G_M$ and a peak width $\sim\Delta{\cal T}^2$.
In Fig.~\ref{fig.position_conductance} we show the differential conductance {\em vs} voltage for different separations between tip and domain wall. As the tip is moved away from the domain wall, nonresonant Andreev reflections become more important and the conductance peak becomes narrower and slightly higher.

\section{Domain wall at the QSHI edge}\label{sec:QSHI_edge}

While the quantized conductance step for the Majorana state derived in the main text is model independent, its robustness to continuum effects was only shown for spinless models described by a $2\times 2$ Hamiltonian.
As an example of a more realistic model, we consider the edge of a quantum spin Hall insulator (QSHI). We calculate the threshold conductance including the continuum contribution for the case when the sample is contacted at the domain wall. We find the universal step at the threshold to be robust in line with the results for the spinless case. This suggests that the conductance step at the threshold may be robust even for more general models of topological superconductors.

The edge of a QSHI, with a domain wall at the origin, can be described by the first quantized Hamiltonian
\begin{gather*}
\mathcal{H}(x)=-iv_F\partial_{x}\sigma_{z}\tau_{z}+\Delta(x)\tau_{x}+B(x)\sigma_{x},\\
\Delta(x)=M-\frac{m(x)}{2},\\
B(x)=M+\frac{m(x)}{2},
\end{gather*}
where
\begin{equation}
m(x)=m\left[\theta(x)-\theta(-x)\right].
\end{equation}
Applying the unitary transformation $U=e^{\frac{i\pi}{4}\sigma_{y}}e^{i\frac{\pi}{4}\tau_{y}}$,
we have
\[
\mathcal{H}\rightarrow\tilde{\mathcal{H}}=U\mathcal{H}U^{\dagger}=\left(\begin{array}{cccc}
M & 0 & 0 & -iv_{F}\partial_{x}\\
0 & -m(x) & -iv_{F}\partial_{x} & 0\\
0 & -iv_{F}\partial_{x} & m(x) & 0\\
-iv_{F}\partial_{x} & 0 & 0 & -M
\end{array}\right).
\]
Let $M\gg m$, and focus on the inner block, which decribes a Dirac field with a domain wall created by a mass jump, as discussed in Sec.~\ref{sec:Majorana_Dirac}.
According to Eq.~(\ref{eq:GF_Dirac}), the Green function at the domain wall is
\begin{equation}
\tilde{g}(0,\omega)=\pi\nu_s\frac{m}{\omega}\left(\begin{array}{cccc}
0 & 0 & 0 & 0\\
0 & \sqrt{1-\frac{\omega^{2}}{m^{2}}} & i & 0\\
0 & -i & \sqrt{1-\frac{\omega^{2}}{m^{2}}} & 0\\
0 & 0 & 0 & 0
\end{array}\right).
\end{equation}
We rotate it back to the original basis $g(0,\omega)=U\tilde{g}(0,\omega)U^{\dagger}$ and obtain the full Green function
\begin{equation}
G(0,\omega)=g(0,\omega)(1-\Sigma(\omega)g(0,\omega))^{-1},
\end{equation}
where the self energy is given by
\begin{equation}
\Sigma(\omega)=t^{2}\diag(g(\omega_{-}),g(\omega_{-}),g(\omega_{+}),g(\omega_{+})).
\end{equation}
Calculation reveals 
\begin{align*}
G^{r,eh}(0,\omega) & =\frac{\pi\nu_sm}{\omega\left[1+t^{4}\pi^{4}\nu_s^{2}(\rho(\omega_{-})+\rho(\omega_{+}))^{2}\right]+2i\pi^{2}\nu_smt^{2}\sqrt{1-\frac{\omega^{2}}{m^{2}}}\left[\rho(\omega_{-})+\rho(\omega_{+})\right]}\\
 & \times\left(\begin{array}{cc}
i & \sqrt{1-\frac{\omega^{2}}{m^{2}}}-i\pi\left(\frac{\pi\nu_s\omega}{m}\right)t^{2}(\rho(\omega_{-})+\rho(\omega_{+}))\\
\sqrt{1-\frac{\omega^{2}}{m^{2}}}-i\pi\left(\frac{\pi\nu_s\omega}{m}\right)t^{2}(\rho(\omega_{-})+\rho(\omega_{+})) & -i
\end{array}\right).
\end{align*}
Denoting $\Gamma_{e/h}=4\pi^{2}\nu_smt^{2}\rho(\omega_{\mp})$ we find
\begin{equation}
\|G^{eh}(0,\omega)\|^{2}=\frac{2(\pi\nu_s)^{2}\left[2m^{2}-\omega^{2}(1-(\Gamma_e+\Gamma_h)^{2}/16m^{2})\right]}{\omega^{2}\left[1-(\Gamma_e+\Gamma_h)^{2}/16m^{2}\right]^2+(\Gamma_e+\Gamma_h)^{2}/4}.
\end{equation}
We readily verify that the denominator is dominated by the second term at the threshold. In particular, we see the continuum contribution in the denominator appears proportional to $\omega^2$ as for the spinless case. Hence, the conductance quantization at $eV=\Delta$ is unaffected by the continuum contribution in this model.

\section{Single-particle current contribution to the conductance}\label{sec:single-particle-current}

In this section we analyze the single-particle tunneling current that can flow in addition to the Andreev current discussed above. A single-particle current is possible if the quasiparticle occupying the bound state can relax to the quasiparticle continuum. Such relaxation can occur, e.g., due to inelastic transitions assisted by phonons or photons.
We concentrate on the Majorana state, although a similar analysis is possible for a trivial Andreev state. We neglect the quasiparticle continuum and include relaxation as a phenomenological parameter, without specifying its microscopic origin. We first derive the single-particle current [given in Eq.~(11) of the main text], from which we then calculate the threshold conductance from single-particle tunneling [Eq.~(12)].

\subsection{Self energy due to relaxation processes}
Let us assume that the local environment of the Majorana state introduces relaxation processes,
induced by phonons or photons, via the self-energy $\Sigma_{\rm ph}$. The substrate Green function can be determined from the Dyson series
\begin{equation}
   g_R = g + g\Sigma_{\rm ph} g + g\Sigma_{\rm ph} g\Sigma_{\rm ph} g +\ldots,
\end{equation}
where $g$ is the Green function without relaxation processes.
We approximate the bare substrate Green function $g$ by the Majorana contribution,
\begin{equation}
     g(\omega) = |\psi_M\rangle \frac{1}{\omega} \langle \psi_M|,
\end{equation}
and project the self-energy onto the Majorana subspace introducing
\begin{equation}   
    \Gamma_{\rm qp}= 2  {\rm Im}\langle \psi_M|\Sigma_{\rm ph}(0)|\psi_M\rangle,
\end{equation}
where we approximate the self-energy by its value at zero energy. The invariance of the Majorana state under particle-hole transformation guarantees that the expectation value $ \langle \psi_M|\Sigma_{\rm ph}(0)|\psi_M\rangle$ is purely imaginary. Thus, the retarded and advanced Green functions of the Majorana state read 
\begin{equation}
    g_{R}^{r,a}(\omega)=\frac{|\psi_M\rangle\langle \psi_M|}{\omega \pm i\Gamma_{qp}/2}.
\end{equation}
In quasi-equilibrium, we can express the greater and lesser Green function in terms of the retarded and advanced Green functions,
\begin{align}
    g^{<}_{R}(\omega)&=f(\omega)(g^a_R(\omega)-g^r_R(\omega))
    =\frac{\Sigma_{\rm ph}^{<}(0)}{\omega^2+\Gamma_{\rm qp}^2/4}
    |\psi_M\rangle\langle \psi_M|   ,\\
    g^{>}_{R}(\omega)&=-(1-f(\omega))(g^a_R(\omega)-g^r_R(\omega))
    =\frac{\Sigma_{\rm ph}^{>}(0)}{\omega^2+\Gamma_{\rm qp}^2/4}
    |\psi_M\rangle\langle \psi_M|,
\end{align}
where $f(\omega)$ is the quasi-equilibrium distribution function and we used the relations
\begin{equation}
    -i\Sigma_{\rm ph}^{<}(0) = \Gamma_{\rm qp} f\,, \qquad   i\Sigma_{\rm ph}^{>}(0) = \Gamma_{\rm qp}(1-f). \label{self-energy}
\end{equation}
These terms are the rates for emptying and filling of the delocalized fermion formed by the Majorana at the contact and a second one far away. Note that local transitions can
change the occupation of this state. Since this fermion has zero energy the two rates are equal $ i\Sigma^{>}_{ph}(0) = -i\Sigma^{<}_{ph}(0)= \Gamma_{\rm qp}/2$ according to detailed balance.  

\subsection{Expressions for the single-particle current}
We can now evaluate the current in Eq.~(\ref{current_formula_green_function}). Besides the Andreev current in Eq.~(\ref{eq:current}), we find the single-particle current from the first term in Eq.~(\ref{eq:GF_l})
\begin{equation}
I_M^s(V)=\frac{e}{4h}\int d\omega
\frac{\Gamma_{\rm qp}[\Gamma_{e}(\omega)n_F(\omega_-)-\Gamma_{h}(\omega)n_F(\omega_+)]-\Gamma_{\rm
qp}[\Gamma_{e}(\omega)(1-n_F(\omega_-))-\Gamma_{h}(\omega)(1-n_F(\omega_+))]}{\omega^{2}+\left(\Gamma_e(\omega)+\Gamma_h(\omega)+\Gamma_{\rm qp}\right)^2/4},\label{Is_app}
\end{equation}
which gives Eq. (11) in the main text if we take  $n_F(\omega_-)\simeq 1$ and $n_F(\omega_+)\simeq 0$ assuming $T \ll \Delta$.

\subsection{Analysis of current}
We compute the single-particle current near the threshold at $eV=\Delta+\eta$ with $\eta \ll \Delta$
\begin{align}
    I_{M}^{s}(V)\simeq&\frac{e}{2h}\int_{-\Delta-\eta}^{\Delta+\eta}d\omega\frac{\omega_{t}^{3/2}\Gamma_{\rm
    qp}\left(\frac{\theta(\eta-\omega)}{\sqrt{\eta-\omega}}+\frac{\theta(\eta+\omega)}{\sqrt{\eta+\omega}}\right)}{\omega^{2}+\left(\frac{\omega_{t}^{3/2}\theta(\eta+\omega)}{\sqrt{\eta+\omega}}+\frac{\omega_{t}^{3/2}\theta(\eta-\omega)}{\sqrt{\eta-\omega}}+\frac{\Gamma_{\rm
    qp}}{2}\right)^{2}}
    \nonumber \\
    &=\frac{e\omega_{t}^{3/2}\Gamma_{\rm qp}}{h}\int_{0}^{\Delta}\frac{d\omega}{\sqrt{\omega}}\frac{1}{(\omega-\eta)^{2}+\left(\frac{\omega_{t}^{3/2}}{\sqrt{\omega}}+\frac{\Gamma}{2}\right)^{2}},
\end{align}
to lowest order in $\eta$. We thus obtain the differential conductance
\begin{equation}
    G_{M}^{s}(x)\simeq\frac{2e^2\Gamma_{\rm
    qp}}{h\omega_{t}}\int_{0}^{\infty}\frac{d\omega}{\sqrt{\omega}}\frac{\omega-x}{\left[(\omega-x)^{2}+\left(\omega^{-1/2}+\Gamma_{\rm qp}/(2\omega_t)\right)^{2}\right]^{2}},
    \quad x=\frac{eV-\Delta}{\omega_t}.
\end{equation}

Now we focus on the conductance at the threshold, namely $x=0$. For weak tip-substrate tunneling, $\omega_{t}\ll \Gamma_{\rm qp}$, 
we find
\begin{align}
    G^s_{M} &\simeq  \frac{2e^2\Gamma_{\rm
    qp}}{h\omega_{t}}\int_{0}^{\infty}d\omega\, \frac{\sqrt{\omega}}{\left[\omega^{2}+\left(\Gamma_{\rm qp}/(2\omega_t)\right)^{2}\right]^{2}}\nonumber \\
    &=\frac{2\pi e^2}{h} \frac{ \omega_t^{3/2}}{\Gamma_{\rm qp}^{3/2}},  
\end{align}
where the $x$-integration is elementary. 
In the opposite limit of strong tip-substrate tunneling, $\omega_{t}\gg \Gamma_{\rm qp}$, we can neglect the contribution of $\Gamma_{\rm qp}$ in the denominator. In this limit, we find
the peak conductance 
\begin{align}
  G^s_{M}(\Delta) &\simeq \frac{2e^2\Gamma_{\rm
    qp}}{h\omega_{t}}\int_{0}^{\infty}d\omega\, \frac{\sqrt{\omega}}{\left[\omega^{2}+1/\omega \right]^{2}}\nonumber \\
    &= \frac{2\pi e^2}{9h} \frac{\Gamma_{\rm qp}}{\omega_t}. 
\end{align}
The parametric dependence of $G^s_M$ is summarized in Eq.~(12) of the main text.


\begin{thebibliography}{99}
\bibitem{foot1} The conductance of Andreev states in $p$-wave or spin-orbit coupled superconductors is discussed in \cite{supp}.
\end{thebibliography}
\end{document}